\documentclass[aps,prl,reprint]{revtex4-2}

\usepackage[utf8]{inputenc}
\usepackage[T1]{fontenc}

\usepackage{amsthm, amsmath, amssymb, amsfonts}
\usepackage[colorlinks=true,allcolors=blue]{hyperref}
\usepackage{xcolor}
\usepackage{listings}

\usepackage{mathtools}
\usepackage{stmaryrd}
\usepackage[colorlinks=true,allcolors=blue]{hyperref}

\newtheorem{theorem}{Theorem}

\newcommand{\x}{\boldsymbol{x}}
\newcommand{\A}{\mathcal{A}}
\newcommand{\B}{\mathcal{B}}
\newcommand{\X}{\mathcal{X}}
\newcommand{\C}{\mathcal{C}}
\newcommand{\nin}{\mathrm{nin}}
\newcommand{\IN}{\mathrm{in}}
\newcommand{\avg}[1]{\langle#1\rangle}
\newcommand{\simp}{\Delta_\X}
\newcommand{\R}{\mathbb{R}}

\global\long\def\Tji{\map(y\vert x)}

\global\long\def\s{\mu}
\global\long\def\N{\mathbb{N}}
\global\long\def\q{q}
\global\long\def\sX{\mathcal{X}}%
\global\long\def\sY{\mathcal{Y}}%

\global\long\def\hlf{{\textstyle \frac{1}{2}}}%
\global\long\def\e{\epsilon}%
\global\long\def\f{f}%
\global\long\def\E{\mathbb{E}}%
\global\long\def\w{q^{\e}}%
\global\long\def\wi{\w_{i}}%
\global\long\def\wip{\left[\M\w\right]_{j}}%
\global\long\def\wp{\M\w}%
\global\long\def\de{{\textstyle \partial_{\e}^{+}}}%
\global\long\def\dem{{\textstyle \partial_{\e}^{-}}}%
\global\long\def\ppi{p(x)}%
\global\long\def\ppl{p(c)}%
\global\long\def\ppll{p^{c}}%
\global\long\def\ppil{\ppll(x)}%
\global\long\def\aei{\ae_{i}}%
\global\long\def\aej{[\map\ae](y)}%
\global\long\def\aei{\ae(x)}%
\global\long\def\aej{\ae(y)}%
\global\long\def\aei{\ae(x)}%
\global\long\def\ae{a^{\e}}%

\newcommand{\sZ}{\mathcal{Z}}

\newcommand{\der}{\frac{\partial }{\partial p_{X_0}(x')}}

\newcommand{\MMC}{\mathrm{MC}}
\newcommand{\D}{\mathrm{D}}
\newcommand{\qm}{\hat{q}}

\newcommand{\rr}{r}
\newcommand{\map}{P}

\newcommand{\DDbase}{D}

\newcommand{\DDf}[2]{\DDbase(#1\Vert #2)}

\newtheorem{applemma}{Lemma}

\newcommand{\LLL}{L}
\lstdefinestyle{mystyle}{
    backgroundcolor=\color{backcolour},   
    commentstyle=\color{codegreen},
    keywordstyle=\color{magenta},
    numberstyle=\tiny\color{codegray},
    stringstyle=\color{codepurple},
    basicstyle=\ttfamily\footnotesize,
    breakatwhitespace=false,         
    breaklines=true,                 
    captionpos=b,                    
    keepspaces=true,                 
    numbers=left,                    
    numbersep=5pt,                  
    showspaces=false,                
    showstringspaces=false,
    showtabs=false,                  
    tabsize=2
}

\begin{document}

\title{Entropy production bounds for systems running computer programs}
%\title{Entropy production of running computer programs}

\author{Abhishek Yadav${}^{ab*}$}
\author{Francesco Caravelli${}^{c}$}
\author{David Wolpert${}^{adef}$}

\affiliation{\vspace{.25cm}${}^a$Santa Fe Institute, 1399 Hyde Park Road
Santa Fe, NM 87501, USA}
\affiliation{${}^b$Department of Physical Sciences, IISER Kolkata, Mohanpur 741246, India}
\affiliation{${}^c$Theoretical Division (T-4), Los Alamos National Laboratory, New Mexico, 87545, USA}
\affiliation{${}^d$Complexity Science Hub, Vienna, Austria}
\affiliation{${}^e$Arizona State University, Tempe, AZ 85281, USA}
\affiliation{${}^f$International Center for Theoretical Physics, Trieste 34151, Italy}
\affiliation{${}^*$Corresponding author: abhi.yadav.28k@gmail.com}

\begin{abstract}
Mismatch cost (MMC) is a universally applicable lower bound on the entropy production (EP) of any fixed physical process
across a given time interval. In the first part of the paper, we establish results concerning MMC to prove that it scales at least linearly with the total heat flow in the worst case over initial distributions. %This establishes that --- in contrast to results like the thermodynamic speed limit theorem or thermodynamic uncertainty relations --- it is often a substantial fraction of the work dissipated on macroscopic scales. 
We also prove that the MMC lower bound over a given time interval never decreases if the time interval is subdivided into a sequence of sub-intervals, and that the bound often increases. In the second part of the paper, we introduce a general framework for computing the minimal EP (i.e., the MMC) associated with running a computer program on any
physical system that implements a modern digital computer. We apply this general framework to compare MMC of running two canonical sorting algorithms, bubble sort and bucket sort. The framework enables us to investigate how thermodynamic cost depends on features like input size and structure (e.g., with or without repeated entries). Finally, we extend the framework to programs that call subroutines.

\end{abstract}

\maketitle
%\tableofcontents 

%\bibliographystyle{apacite}
%\bibliography{example}
%Consider the RASP:
%https://en.wikipedia.org/wiki/Random-access_stored-program_machine

%\textcolor{red}{ADD comment in introduction about }

\section*{Significant statement}
While computational complexity traditionally focuses on time and memory resources, the energetic cost of computation remains poorly understood. This work introduces a general framework to quantify the fundamental energetic cost of running any computer program, opening a new approach to comparing algorithmic efficiency beyond time and memory. Leveraging stochastic thermodynamics, we derive universal lower bounds on the unavoidable energy dissipation incurred by any high-level program. Grounded in stored-program architecture, these bounds are largely independent of physical implementation. Our approach demonstrates how thermodynamic efficiency varies across algorithms, exemplified through sorting algorithms. This framework offers new insights into the physical limits of algorithmic efficiency and energy consumption in computing.

\section{Introduction}
\label{sec:intro}

\subsection{Background}

Computer programs are sequences of instructions that, when executed mechanically, produce a desired computation. In the early days of computing, the very notion of a programmable machine was far from obvious. In his seminal 1936 paper, Alan Turing rigorously formalized the concept of a universal computer---a machine capable of executing any sequence of instructions by storing them on a tape or in memory \cite{turing1936computable}. Building on this foundation, the subsequent development of stored-program machines, later known as the von Neumann architecture, enabled the proliferation of modern computers. This architecture also gave rise to the formal study of program complexity, focusing on the resources---most notably time and memory---required to compute a function. These ``time'' and ``space'' costs became central measures of program efficiency.

An analogous study of the energetic cost of executing a computer program has remained largely unexplored---despite its apparent importance for real-world computing---due to several challenges. First, classical thermodynamics and equilibrium statistical mechanics are ill-suited for machines operating far from equilibrium with many interacting, dynamically evolving degrees of freedom. Second, real-world computational machines are built on a wide variety of physical substrates, making it difficult to formulate a universal framework applicable to any computer program. In this paper, we address these challenges by adopting tools from stochastic thermodynamics and applying the mismatch cost (MMC) lower bound on entropy production to a logically abstract stored-program model of computation. This is the same foundational model that enabled early computer science to compare the time and space complexity of programs in a unified manner, and it likewise allows us to derive lower bounds on thermodynamic cost of computer programs. 

To provide background on the first major challenge, earlier foundational work by Szilard~\cite{szilard1964decrease}, Landauer~\cite{landauer1961irreversibility}, and Bennett~\cite{bennett1982thermodynamics} established that logically irreversible computations necessarily produce a minimum amount of heat, quantified as $k_B T \Delta S$, where $k_B$ is Boltzmann’s constant, $T$ is the temperature of the surrounding reservoir, and $\Delta S$ is the change in the system’s Shannon entropy. This reasoning led to the widely held belief that logically reversible computations, for which $\Delta S = 0$, have no fundamental energetic cost~\cite{sagawa2014thermodynamic}. Moreover, the entropy change $\Delta S$ accounts only for the reversible portion of heat exchange—heat that, in principle, can be fully recovered by reversing the process~\cite{van2015ensemble}. Consequently, this framework does not capture the irreversible energy dissipation arising from the far-from-equilibrium dynamics typical of real computational machines.

Prior studies on the energetic efficiency of computer programs have largely relied on Landauer’s principle~\cite{bingham2011modeling, tyagi2016toward}, attributing an energy cost of $k_B T \ln 2$ to each bit erasure step. However, these approaches neglect the energetic cost of logically reversible operations and fail to capture the irreversible dissipation arising from the far-from-equilibrium dynamics inherent in program execution. 

It is now well established that, beyond the Landauer cost, there exists an inherently irreversible component known as entropy production (EP). EP quantifies the energy irreversibly dissipated into the environment~\cite{van2015ensemble, seifert2012stochastic}, and it can be strictly positive even during logically reversible computations. This deeper understanding arises from major advances in non-equilibrium statistical physics, which have extended classical thermodynamics to systems operating at mesoscopic scales and far from equilibrium. In particular, the framework of stochastic thermodynamics~\cite{seifert2012stochastic, parrondo2015thermodynamics, wolpert2019stochastic} has proven invaluable for characterizing far-from-equilibrium behavior and quantifying energy dissipation across a wide range of processes.  As such, it provides a rigorous and well-suited theoretical foundation for studying the energetic costs of computation~\cite{wolpert2024stochastic}.

One important contribution to total EP is the mismatch cost (MMC)~\cite{kolchinsky2017dependence, kolchinsky2021dependence}. MMC quantifies the extra entropy production that arises when the starting probability distribution of a system differs from the optimal distribution that minimizes EP. Thus, even processes that produce zero EP for a specific initial distribution—making them thermodynamically reversible under those conditions—can generate positive EP precisely equal to the MMC if the initial distribution changes.
This concept has been further explored to show that computational tasks that are inherently modular---that is, when a system is decomposed into various sub-parts for computation---have an unavoidable thermodynamic cost associated with them~\cite{wolpert2020thermodynamics}, establishing fundamental bounds on entropy production in systems such as communication channels~\cite{yadav2025minimal} and Boolean circuits~\cite{yadav2025circuits}. 

\subsection{Contributions}

The contributions of this paper are twofold. First, we establish new theoretical results concerning MMC in Sec.~\ref{sec_MMC}. Specifically, we show that the MMC contribution to the total thermodynamic cost grows at least linearly with the total heat flow in the worst case over the initial distribution. This establishes that --- in contrast to results like the thermodynamic speed limit theorem or thermodynamic uncertainty relations --- MMC in principle could constitute a substantial fraction of the total dissipated heat on macroscopic scales. Additionally, in Sec.~\ref{subsec_coarse_graning} we prove that the sum of MMCs evaluated at finer time resolutions always exceeds the MMC computed over a coarser-grained execution that omits intermediate steps---thus characterizing MMC’s behavior under time coarse-graining. This implies that MMC remains a valid lower bound on total EP even when considering only higher-level computational steps without resolving the detailed physical dynamics. Moreover, the sum of MMCs evaluated on any subset of computational steps still lower-bounds the total MMC, and thus total EP, over the entire process.

Second, we evaluate the MMC associated with executing computer programs. In Sec~\ref{sec_RASP}, we introduce the framework based on the stored-program architecture of modern computers where a program counter keeps track of current instruction and a clocked, iterative process sequentially modifies memory contents. 
This framework not only allows us to define a computer program’s full state---including all variable values and the program counter---at any point during its execution but also to determine the MMC for each iteration of the machine. It thereby provides a minimal thermodynamic cost that is incurred in each step of a program. %Specifically, we use a model grounded in the principles of stored-program architecture, to formally define and track a program’s state during execution. This model allows us to define a program’s full state---including all variable values and the program counter---at any point during execution.

We then apply this framework to a range of concrete examples, including classic sorting algorithms. Furthermore, we extend the analysis to encompass programs that invoke subroutines, thereby illustrating how MMC can be used to quantify the thermodynamic costs of modular, hierarchical program structures.

% In Sec.~\ref{subsec_stored_program}, we briefly review stored-program computers, emphasizing their periodic operation and the critical role of the program counter in sequencing instructions. In Sec.~\ref{subsec_RASP}, we introduce the Random Access Stored Program (RASP) model, which abstracts away hardware-specific details while retaining core architectural elements such as registers, memory, and instruction sequencing. This model allows us to define a program’s full state---including all variable values and the program counter---at any point during execution. The set of all such states forms the program’s state space. For any given input distribution, the program traces a sequence of states that can be represented as a directed graph, whose adjacency matrix encodes the full execution flow. This representation enables a compact description of the discrete-time dynamics of the probability distribution over the program’s states at each step (see Eq.~\ref{eq_G_evolution}).

% In Sec.~\ref{sec_MMCprograms}, we apply this framework to several computer programs, computing both the per-step and total MMC during their execution. We analyze bubble sort in detail, exploring how its state space and MMC evolve as we vary input size and structure. We compare cases where inputs are restricted to permutations versus those allowing repeated entries, illustrating how input characteristics impact the MMC.

% Finally, Sec.~\ref{sec_conclusion} summarizes our main findings and discusses potential future research directions.

\section{General properties of the mismatch cost}\label{sec_MMC}

% Equilibrium statistical mechanics derives thermodynamics properties of a system from the function of the underlying equilibrium statistics. Stochastic thermodynamics is an extension which defines thermodynamic quantities such as heat, work, and entropy to the level stochastic trajectories in processes evolving arbitrarily far-from-equilibrium. The key assumption in stochastic thermodynamics is that any degree of freedom not explicitly described by the system's dynamics—such the heat or particle reservoirs remains in equilibrium. This allows us to connect thermodynamic quantities and dynamics, through the principle of local detailed balance~\cite{seifert2012stochastic, seifert2018stochastic, esposito2010entropy}. 

{\color{black} We begin with some general considerations from stochastic thermodynamics and the role of the MMC. We consider a system with a state space $\X$ that undergoes a transformation. Starting from an initial state $x_0 \in \X$, the system evolves and ends in a state $x_1 \in \mathcal{X}$. This transformation can be described by a conditional map $G(x_1 | x_0)$, which specifies the probability that the system ends in state $x_1$ given that it started in state $x_0$.

This formulation captures a wide range of processes, including computational devices such as Boolean gates (where the system state changes as a gate computes an output), Boolean circuits executing sequences of operations, and chemical reaction networks where the composition of reactants change over time. In general, the mapping from an initial state to a final state is stochastic rather than deterministic, meaning that $G(x_1 | x_0)$ typically takes values strictly between 0 and 1.}

If the initial state is drawn from a distribution $p_{X_0}$ over $\mathcal{X}$, then the final state of the system is distributed according to $p_{X_1}$ which is given by,
\begin{equation}
    p_{X_1}(x_1) = \sum_{x_0\in \X}G(x_1|x_0)p_{X_0}(x_0)
\end{equation}
As a short-hand we sometime write $p_{X_1} = Gp_{X_0}$. Let $\simp$ denote the probability simplex associated with the state space $\X$, i.e., the set of all probability distributions over system states. 

Associated with any such transformation $p_{X_0} \to p_{X_1}$, many thermodynamic cost functions take the common mathematical form~\cite{kolchinsky2021dependence}:
\begin{equation}\label{EPdef}
    \C(p_{X_0}) = S(Gp_{X_0}) - S(p_{X_0}) + \sum_{x \in \X} f(x) p_{X_0}(x),
\end{equation}
where $f$ is a given real-valued function on the state space $f:\X \to \R$, and $S(p)$ is the Shannon entropy defined as $S(p) = - \sum_{x\in \X} p(x) \log p(x)$. Depending on the physical interpretation of $f$ the cost function \eqref{EPdef} can represent a variety of thermodynamic quantities, including total EP, non-adiabatic EP~\cite{esposito2010three}, free energy loss~\cite{kolchinsky2025maximizing}, and entropy gain~\cite{plastino1995fisher}. In particular, when $f(x)$ represents the average heat flow during the process when the system starts in state $x$, then~(\ref{EPdef}) corresponds to the dissipated work or the EP. 

{\color{black} The values $f(x)$ associated with thermodynamic costs can, in principle, be either measured experimentally or computed from detailed knowledge of the underlying physical dynamics. Once $f(x)$ is known, the thermodynamic cost \eqref{EPdef} is fully determined for any initial distribution. In practice, however, neither the microscopic degrees of freedom relevant for measuring $f(x)$ nor a complete description of the system’s dynamics are typically accessible. As a result, determining the thermodynamic cost associated with a given computational map $G$ is generally a challenging task.

Much of the development of stochastic thermodynamics has therefore focused on identifying and bounding specific contributions to thermodynamic cost that arise from physical constraints on the process. Examples include thermodynamic uncertainty relations~\cite{barato2015thermodynamic, gingrich2016dissipation, horowitz2020thermodynamic}, which quantify costs associated with precision of currents, and speed limit theorems~\cite{shiraishi2018speed, vo2020unified}, which capture costs imposed by finite-time operation. Such results usually rely on simplifying assumptions about the underlying dynamics, most commonly Markovianity, weak coupling, or detailed balance.

In contrast, thermodynamics of computation is often driven by a question of a more universal kind: what thermodynamic cost is unavoidably incurred in implementing a given map $G$, independent of the detailed physical realization? MMC has been developed and applied across a wide range of computational settings---including finite automata~\cite{ouldridge2023thermodynamics}, Turing machines~\cite{TuringKolchinskyWolpert2020}, communication systems\cite{yadav2025minimal}, and logical circuits~\cite{yadav2025circuits}---and provides bounds that are largely independent of the specific physics of the underlying implementation. Below, we review the MMC framework and summarize several of its key properties.}

For any fixed physical process implementing the stochastic map $G$, which transforms any initial distribution $p_{X_0}$ into a final distribution $G p_{X_0}$, there exists among all possible initial distributions an optimal distribution $q_{X_0}$ that minimizes the thermodynamic cost,
\begin{equation}\label{Prior_def}
q_{X_0} = \arg\min_{r\in \Delta_{\X}} \C(r),
\end{equation}
where $q_{X_0}$ has full support over state space $\X$ as long as the map $G$ is not a deterministic map~\cite{kolchinsky2017dependence}. 

When the same process runs for an initial distribution different from $q_{X_0}$, the process incurs an extra cost in addition to that incurred by the optimal initial distribution $q_{X_0}$ which minimizes that cost. The total cost for any $p_{X_0}$ decomposes into~\cite{kolchinsky2017dependence, wolpert2020thermodynamics, kolchinsky2021dependence}:
\begin{equation}\label{eq:MMC1}
    \C(p_{X_0}) = D(p_{X_0}||q_{X_0}) -  D(Gp_{X_0}||Gq_{X_0})  + \C(q_{X_0})
\end{equation}
%Consequently, the decomposition in Eq.~\ref{eq:MMC1} applies equally to all such thermodynamic cost functions.
where $D(\cdot \| \cdot)$ denotes the Kullback–Leibler (KL) divergence. 

The optimal distribution $q_{X_0}$ is also referred to as the {\it prior distribution} associated with the cost function $\C$. The decomposition in Eq.~(\ref{eq:MMC1}) expresses the total cost as the sum of two terms: the minimum achievable cost $\C(q_{X_0})$, and an additional term arising from the mismatch between any actual distribution $p_{X_0} \in \simp$ and the prior $q_{X_0}$. This additional term, given by the drop in KL divergence under the map $G$, is known as the mismatch cost:

\begin{equation}\label{eq:MMCdef}
    \MMC(p_{X_0}) = D(p_{X_0} \| q_{X_0}) - D( Gp_{X_0} \| Gq_{X_0}).
\end{equation}
Due to the data-processing inequality for KL-divergence, MMC is always non-negative: $\MMC(p_{X_0}) \ge 0$ for any $p_{X_0}$~\cite{polyanskiy2014lecture}. The formula for MMC is very general, applicable to classical systems, quantum systems,and even systems undergoing non-Markovian dynamics. The formula for MMC applies broadly whether the dynamics is Markovian or non-Markovian process, classical or quantum, and whether it is in discrete or continuous time. {\color{black} The prior distribution defined in Eq.~\eqref{Prior_def}  encodes features of the process through which $G$ is realized. In this sense, it is important to emphasize that the MMC~(\ref{eq:MMCdef}) is not completely independent of the underlying physical implementation.}

{\color{black} That said, in many settings some natural constraints can easily be reflected in the properties of the prior distribution. For example, if the system undergoing the transformation consists of two or more physically independent subsystems, then any physically admissible prior over the joint system must factorize as a product distribution over those subsystems~\cite{wolpert2019stochastic}. As a consequence, for a generic initial distribution the system incurs an unavoidable EP, captured by the MMC. In this sense, the resulting bounds are universal: they do not require a detailed dynamical description of the underlying physical implementation. For instance, if two Boolean gates in a circuit, each running on its own underlying physical process that is separate from the other, operate on inputs that are correlated, it results in a thermodynamic cost that can be quantified by the MMC, independent of the microscopic physics of the gates~\cite{yadav2025minimal}. In general, computational devices are designed based on some modular and hierarchical design principles, providing strong constraints on how their subsystems are connected.}

Another set of examples where MMC provides strictly positive contribution to the thermodynamic cost is when a process is repeated over and over, such as in digital computers that run periodic processes governed by a global clock. Suppose a process that is characterized by the associated  map $G$ is repeatedly applied to distribution over states of the system, without reinitializing the system. That is, starting from $p_{X_0}$ at time $t = 0$, the system evolves through a sequence of distributions $\{p_{X_0}, p_{X_1}, \dots, p_{X_n}\}$, where each $p_{X_{t+1}} = G p_{X_t}$. Although the actual state distribution changes over time, the underlying process implementing $G$—and hence the associated prior distribution $q_{X_0}$—remains fixed across all iterations. The total MMC then accumulates over iterations as:

\begin{align}
    \MMC(p_{X_0}) &= \sum_{i = 0}^{n-1} \D(p_{X_i}||q_{X_0}) - \D(p_{X_{i+1}}||q_{X_1}) \\
    &= \sum_{i = 0}^{n-1} \D(G^{i}p_{X_0}||q_{X_0}) - \D(G^{i+1}p_{X_0}||Gq_{X_0}) \label{eq:PMMC_def}
\end{align}
Note that even if the process starts at the prior $q_{X_0}$, after the first iteration, the distribution becomes $p_1 = G q_{X_0} $, which differs from $q_{X_0}$. This deviation from the prior distribution results in a strictly positive MMC in the next iteration, and the same holds for subsequent iterations~\cite{ouldridge2023thermodynamics}. 

The prior distribution defined in~(\ref{Prior_def}) is determined by the cost function $f$ and the stochastic map $G$. In the following section, we examine how specific properties of $f$ influence the properties of the prior, and consequently the MMC. In particular, we show that the maximum and minimum values of $f(x)$ across states $x \in \X$, especially in the regime where $f(x)$ is large, provide a lower bound on the MMC’s contribution to the total cost.

\subsection{Prior distribution}\label{subsec_prior}

Consider a scenario where
% where the total heat generation is large on a scale of $\ln|\X|$, i.e., when $f(x) \gg \ln|\X|$, and $f(x)$ is not uniform across all states $x \in \X$, the resulting prior distribution  gets closer to the edges of the probability simplex. In other words, the prior becomes increasingly peaked. This peaking of the prior results in a high value of KL divergence between a typical actual distribution $p_{X_0}$ and the prior $q_{X_0}$, thereby increasing the MMC. % To sketch the intuition, 
$f(x)$ values are sufficiently large for all $x$ such that the right hand side of Eq.~(\ref{EPdef}) is dominated by $\sum_{x \in \X} f(x) p_{X_0}(x)$ to large extent compared to the term $S(p_{X_0}) - S(Gp_{X_0})$. In such cases, if $f(x)$ is not uniform across all states---meaning $f(x)$ is higher for some states than others---the associated prior distribution $q_{X_0}$ will take on lower values for those higher-weighted states compared to the rest. The greater the non-uniformity of $f(x)$ across states, the closer the prior will be to the edge of the simplex $\simp$. Then, any typical actual distribution $p_{X_0}$ on the simplex that is not close to the edge would yield a significantly high value of KL-divergence $D(p_{X_0}\|q_{X_0})$. This intuitive idea is formalized in App.~\ref{app:mismmin} to prove that in the worst case over initial distributions, MMC scales at least linearly with the difference between maximum and minimum value of $f(x)$,

\begin{equation}\label{eq:worst_case_MMC}
\MMC^* \ge \max_x\{f(x)\} - \min_x\{f(x)\} - \log|\X|. 
\end{equation}

In this equation, $\MMC^*$ is the MMC for the initial distribution that is furthest from the prior distribution in terms of KL divergence, and therefore incurs the maximum MMC. As an example, consider the case where $f(x)$ represents the average heat flow into the environment when the system starts in state x. At sufficiently large scales---particularly in the regime where computational processes operate---these heat flow values are on the order of $(k_B T)^{-1} \sim 10^{23}$, while the size of the state space, $|\X|$, is much smaller, around $10^2$. As a result, the difference $\max_x{f(x)} - \min_x{f(x)}$ is typically of the same order, $(k_B T)^{-1} \sim 10^{23}$. This implies that, in the worst-case scenario, the MMC contribution to the total cost---given by~(\ref{eq:worst_case_MMC})---can be comparable to the total cost itself. While many EP bounds, such as the thermodynamic uncertainty relation and speed limit theorems, offer meaningful insights at the microscopic level, their relevance diminishes at mesoscopic or macroscopic scales. In these regimes, where heat generation is substantial and relatively easy to measure experimentally, such bounds capture only a small fraction of the total EP. In contrast, Eq.~(\ref{eq:worst_case_MMC}) shows that when $f(x)$ takes large values---corresponding to macroscopic processes---the MMC can contribute significantly to the overall entropy production.

\subsection{Time coarse-graining}\label{subsec_coarse_graning}

Consider a system with state space $\X$, and let $X_0$, $X_1$, and $X_2$ denote the random variables representing the system state at three successive time steps $t_0$, $t_1$, and $t_2$, with corresponding distributions $p_{X_0}$, $p_{X_1}$, and $p_{X_2}$. Due to the additivity of the cost function over time~\cite{parrondo2009entropy}, we have:
\begin{equation}\label{eq_tcg1}
    \C_{02}(p_{X_0}) = \C_{01}(p_{X_0}) + \C_{12}(p_{X_1}),
\end{equation}
where $\C_{ij}$ denotes the cost associated with the transition from time $t_i$ to $t_j$.

Let $q_{X_0}$ be the prior distribution that minimizes $\C_{02}$, and let $R_{02} = \C_{02}(q_{X_0})$ denote the corresponding residual cost. Since $q_{X_0}$ is not necessarily the prior for $\C_{01}$, we obtain the following decompose of $\C_{01}$:
\begin{equation}\label{eq_tcg2}
    \C_{01}(q_{X_0}) = \MMC_{01}(q_{X_0}) + R_{01},
\end{equation}
where $R_{01}$ is the residual cost for $\C_{01}$ and $\MMC_{01}(q_{X_0})$ is the mismatch cost of using $q_{X_0}$ instead of the prior for $\C_{01}$.

Let $\qm_{X_1}$ denote the distribution at time $t_1$ obtained by evolving $q_{X_0}$ under the dynamics from $t_0$ to $t_1$. This intermediate distribution is not generally the prior for $\C_{12}$, so using the MMC decomposition of $\C_{12}$, we get:
\begin{equation}\label{eq_tcg3}
    \C_{12}(\qm_{X_1}) = \MMC_{12}(\qm_{X_1}) + R_{12}.
\end{equation}
where $R_{12}$ is the residual cost of dynamics from time $t_1$ to $t_2$. The residual cost of the full dynamics from time $t_0$ to $t_2$ is by definition $R_{02} = C_{02}(q_{X_0})$. 

Using equations~(\ref{eq_tcg1}),~(\ref{eq_tcg2}),~(\ref{eq_tcg3}), we get,
\begin{align}
    R_{02} &= \C_{02}(q_{X_0}) \nonumber \\ 
           &= \C_{01}(q_{X_0}) + \C_{12}(\qm_{X_1}) \nonumber  \\
           &= \MMC_{01}(q_{X_0}) + R_{01} + \MMC_{12}(\qm_{X_1}) + R_{12} \nonumber  \\
           &\ge R_{01} + R_{12}. \label{eq_tcg4}
\end{align}
where in the last inequality we have used the non-negativity of the MMCs $\MMC_{01}$ and $\MMC_{12}$. For any arbitrary initial distribution $p_{X_0}\in \simp$, using the MMC decomposition in Eq.~(\ref{eq_tcg1}) results in,
\begin{equation}
    \MMC_{02}(p_{X_0}) + R_{02} = \MMC_{01}(p_{X_0}) + R_{01} + \MMC_{12}(p_{X_1}) + R_{12}
\end{equation}
By using, $R_{02}\ge R_{01} + R_{12}$, from~(\ref{eq_tcg4}), we obtain the desired result:
\begin{equation}
    \MMC_{02}(p_{X_0}) \le \MMC_{01}(p_{X_0}) + \MMC_{12}(p_{X_1}).
\end{equation}
for any $p_{X_0} \in \simp$. Therefore, the MMC computed over the full interval $[t_0, t_2]$—without accounting for the intermediate step—is less than or equal to the sum of the MMCs computed separately over $[t_0, t_1]$ and $[t_1, t_2]$. This demonstrates that the MMC at a coarser temporal resolution is always bounded above by the total MMC at a finer resolution over the same time span.

An analogous inequality does not hold for spatial coarse-graining. Let $(X, Y)$ denote a pair of random variables describing the system at a finer spatial resolution, and let $X$ alone represent the coarse-grained description. The cost function at the fine resolution is given by:
\begin{equation}
C_{XY}(p_{XY_0}) = \avg{f_{XY}}_{p_{XY_0}} + S(p_{XY_1}) - S(p_{XY_0}).
\end{equation}
By defining $f_X(x) = \sum_y p_{Y|X}(y|x) f_{XY}(x, y)$ and marginalizing over the variable $Y$, one can derive the corresponding coarse-grained cost,
\begin{equation}
\C_X(p_{X_0}) = \avg{f_X}_{p_{X_0}} + S(p_{X_1}) - S(p_{X_0}).
\end{equation}
Note that $f_X$ is dependent on $p_{Y|X}$ and therefore  on $p_{XY}$. While $\MMC_{XY}(p_{XY})$ is well-defined over the entire space $\Delta_{\mathcal{XY}}$, the coarse-grained mismatch cost $\MMC_X(p_X)$ is only well-defined if the function $f_X$ is uniquely determined. However, since $f_X$ varies with $p_{XY}$, $\MMC_X(p_X)$ lacks a consistent definition and therefore a direct comparison between $\MMC_{XY}(p_{XY})$ and $\MMC_X(p_X)$ is ill-posed.

In the following sections of the paper, we introduce the foundational stored-program architecture of modern computers and use the MMC framework to analyze the thermodynamic cost of running a computer program on such a machine. {\color{black} Importantly, the treatment is more general: it uses only the MMC expression in Eq.~(\ref{eq:MMCdef}) together with the notion of repeated physical processes and the associated MMC expression in Eq.~(\ref{eq:PMMC_def}) and does not require non-uniformity of $f(x)$. It does not rely on the results concerning linear lower bounds on MMC or the time coarse-graining result discussed in Sec.~\ref{subsec_prior} and~\ref{subsec_coarse_graning} respectively. }

\begin{figure*}
    \centering
    \includegraphics[trim = {0 10cm 0 0}, width=\linewidth]{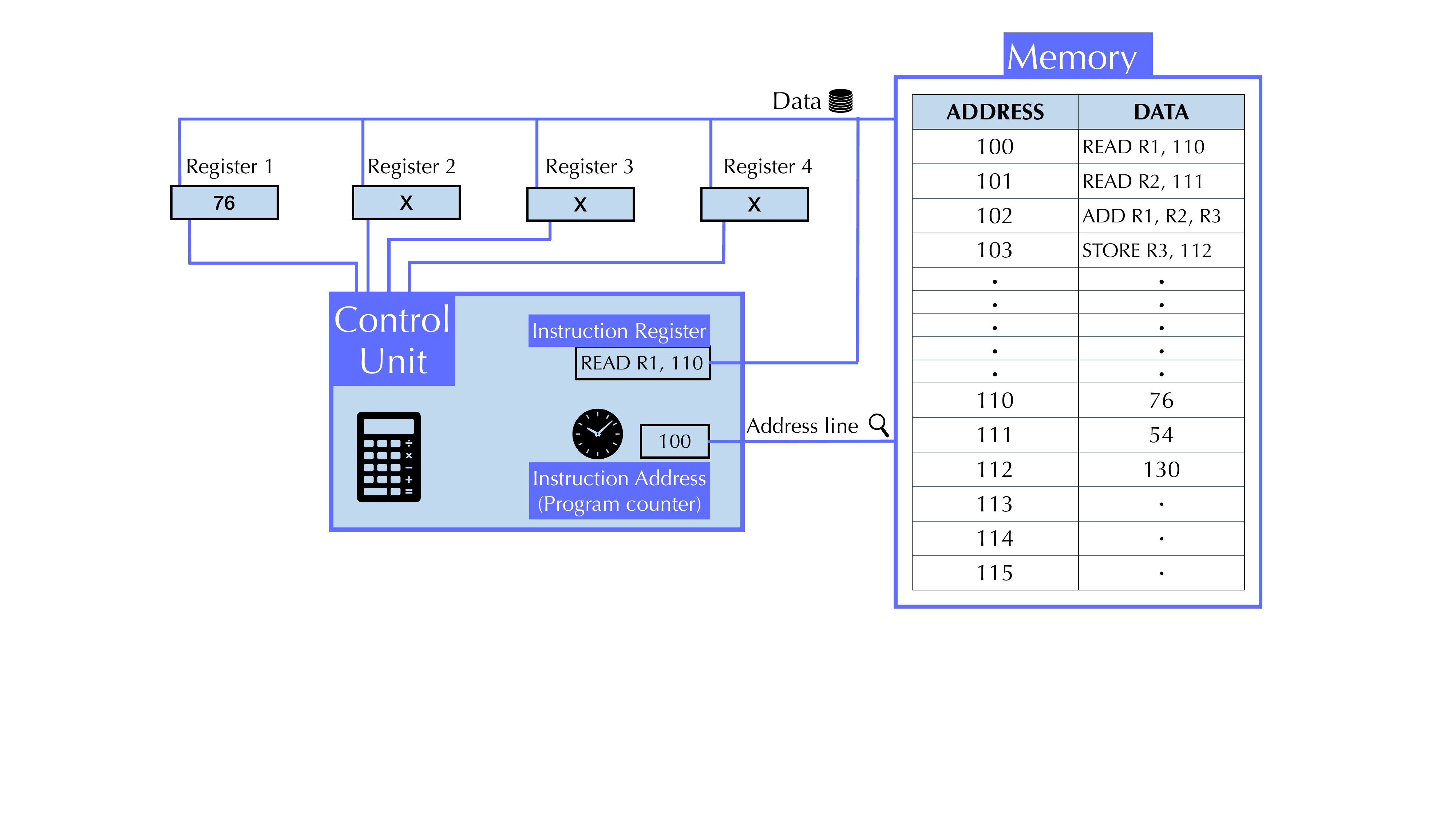}
    \caption{Stored program architecture: The control unit has access to multiple registers, including the instruction register and program counter. Memory stores the data and instruction which can be accessed using address. Control unit communicates with memory using the address and data line, aside from read and write enable lines (not shown in the figure). PC holds the address of current instruction, which is used to fetch the instruction into IR. With the help of check circuits (not shown in the figure), the instruction is decoded into control signals that change register activity, ALU configuration, and memory access. With each clock cycle, the program counter increments and points to the next instruction.}
    \label{fig1}
\end{figure*}

\section{Framework}\label{sec_RASP}
%\subsection{Fundamentals of stored-programs and modern computer architecture}\label{sec_RASP}

% How can a single piece of hardware---the CPU---execute so many different instructions without being rewired for each new task? 

\subsection{Stored program computer}\label{subsec_stored_program}

A defining feature of modern computers is their programmability -- a single machine capable of executing arbitrary well-defined set of instructions. %Before modern computers, ``computation'' was done either by people or by specialized machines. Human computers followed long, fixed sequences of instructions by hand, so changing the task simply meant giving them a different list of instructions. This made them highly programmable, but extremely slow. 
{\color{black} Early computing machines though considered orders of magnitude faster then human computer, were not easily programmable. To perform a new task, engineers often needed to rewire the device, flip physical switches, or redesign parts of the hardware. In effect, a different computation meant a different machine. The stored-program architecture was a conceptual leap because it introduced the idea that instructions could be stored in memory in exactly the same formal representation as data, while execution is handled by a separate processing unit. Programming no longer required hardware modification; it meant writing a new list of instructions to memory --- programming as we know it in the modern sense, though at the machine level.}

This architecture---now standard in digital computers---enabled programmability and ultimately shaped modern computing. In essence, a stored-program computer consists of a memory that holds both data and instructions, and a control/processing unit that fetches instructions by address, updates a program counter, and executes those instructions on the stored data.

The control unit includes several small storage locations, called registers, that hold data or instructions during processing, two of which are central to program execution: The instruction address register, also called the program counter (PC), which stores the memory address of the current instruction; and the instruction register (IR), which holds the current instruction fetched from memory. This register is also referred to as the program counter. The control unit also contains an arithmetic logic unit (ALU) responsible for performing basic arithmetic and logical operations (see Fig.~\ref{fig1}). 

Fig.~\ref{fig1} provides a stylized sketch of this architecture with an example. In Fig.~\ref{fig1} the program counter initially points to memory address 100. The control unit, using address 100 loads the instruction stored at that location into the IR. Suppose the instruction is \texttt{`READ R1, 110'} (encoded in binary in memory), which means load the value from memory address 110 into register R1. This instruction enables register R1 for writing and issues a read request to memory address 110. For an instruction like \texttt{`ADD R1, R2, R3'}, which adds the values in R1 and R2 and stores the result in R3, the check circuits direct the ALU to accept inputs from R1 and R2, perform an addition, and write the result to R3.

At the end of each instruction, the program counter increments, advancing to the address of the next instruction, and the cycle repeats. The clock-driven update of the program counter is what defines the control flow, and any deviation (e.g., jumps or branches) must be explicitly encoded in the instruction sequence.

This process, called the fetch-execute cycle, repeats with every clock cycle, using the same physical logic circuitry, regardless of which instruction is being executed. The control unit is not reconfigured or altered between tasks; instead, it behaves as a periodic dynamical system, transitioning deterministically based on the current instruction and register states. This uniformity across different instructions---performing the same physical sequence over and over while producing different outcomes depending on the instruction---is a defining characteristic of stored-program machines.

\begin{figure*}
    \centering
    \includegraphics[trim = {0 7cm 0 0}, width=0.9\linewidth]{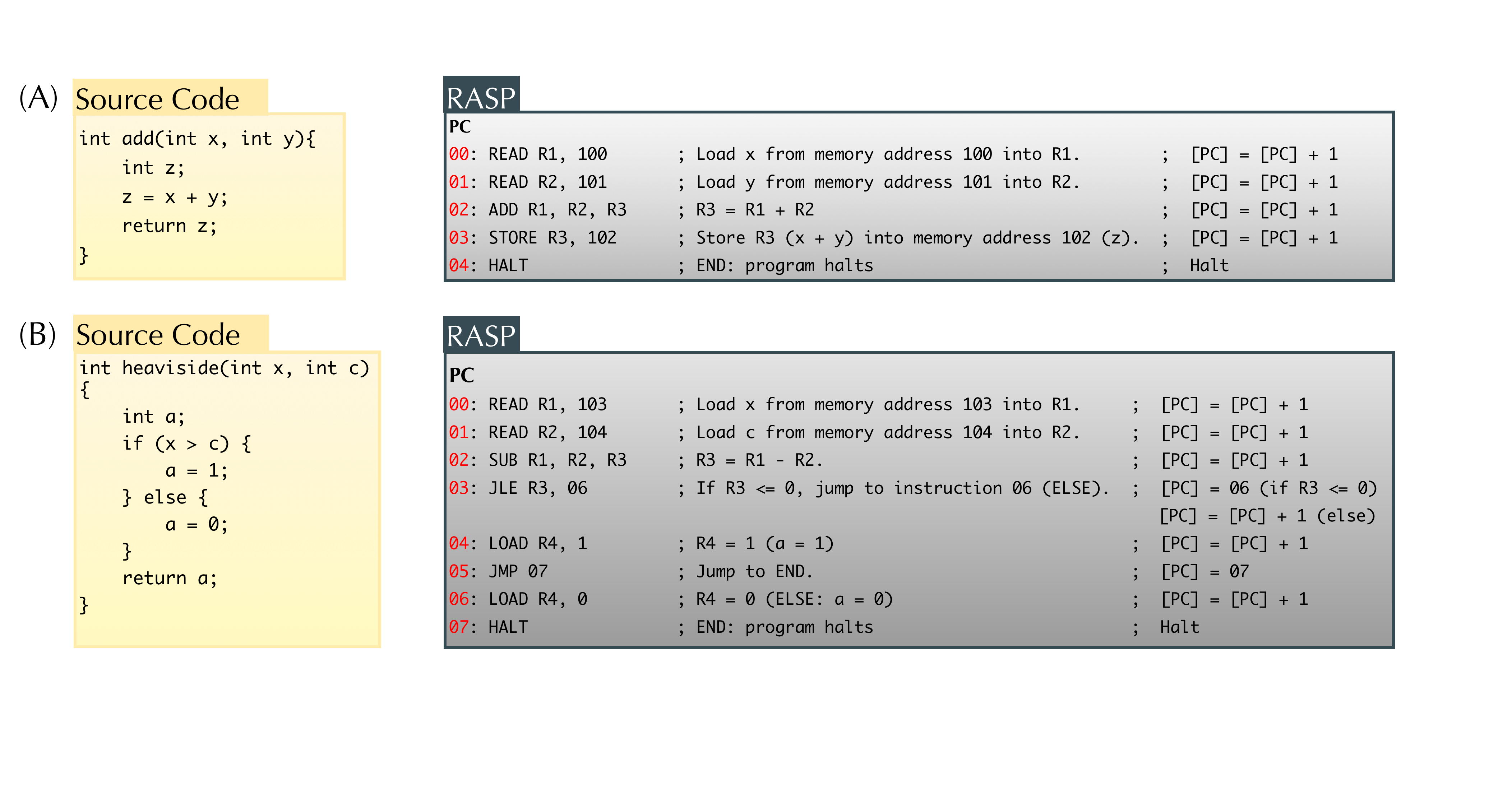}
    \caption{Two C programs are translated into their corresponding lower-level RASP representations. The RASP code explicitly shows how each variable is assigned to a register and how the program counter advances with each instruction. In Program (a), which performs simple addition, the program counter increments sequentially since there are no loops or conditionals. Each instruction—such as loading values and performing arithmetic—corresponds to one step. In contrast, Program (b) includes conditional logic, which introduces non-sequential control flow. Here, the program counter may jump to a different instruction depending on the outcome of a conditional check. A table outlining the meaning of the commands used in the RASP language is provided in Table~\ref{TABLE_RASP} in Appendix.}
    \label{fig2}
\end{figure*}

\subsection{Modeling program dynamics using RASP}\label{subsec_RASP}

\begin{figure}
    \centering
    \includegraphics[width=0.7\linewidth]
    {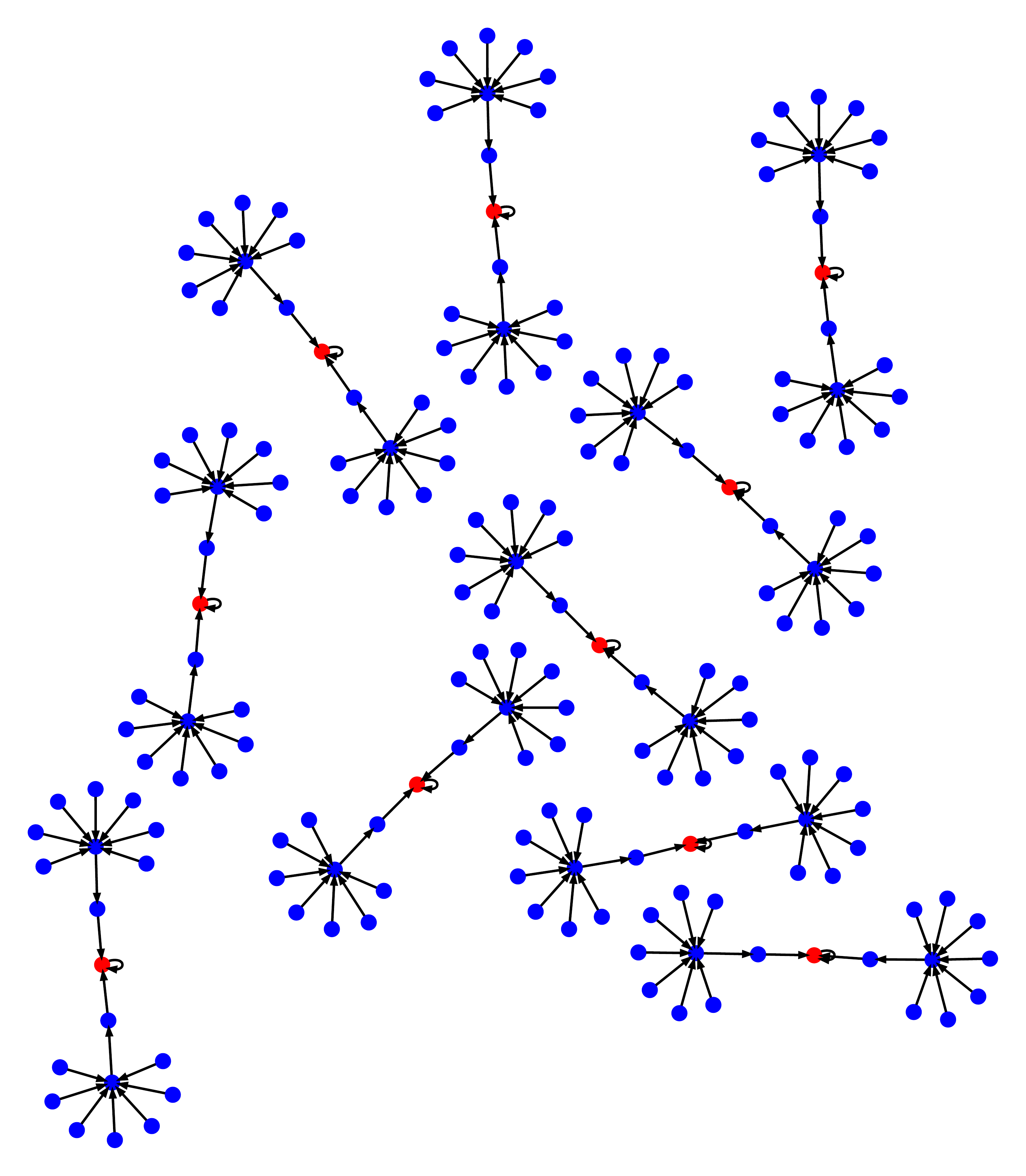}
    \caption{State space of the \texttt{heaviside} program execution defined in Fig.~\ref{fig2}. Each node represents a unique program state, defined by the values of all registers and the program counter, generated for input values $c = 5$ and $x \in \{0, 1, \dots, 9\}$. Leaf nodes (with no incoming edges) correspond to the program’s possible initial states. Directed edges trace the sequence of state transitions during execution. All paths eventually converge to a halting state (red), where the program halts.}
    \label{fig_heaviside_statespace}
\end{figure}

\begin{figure*}
    \centering
    \includegraphics[trim = {0 12cm 0 0}, width=0.9\linewidth]{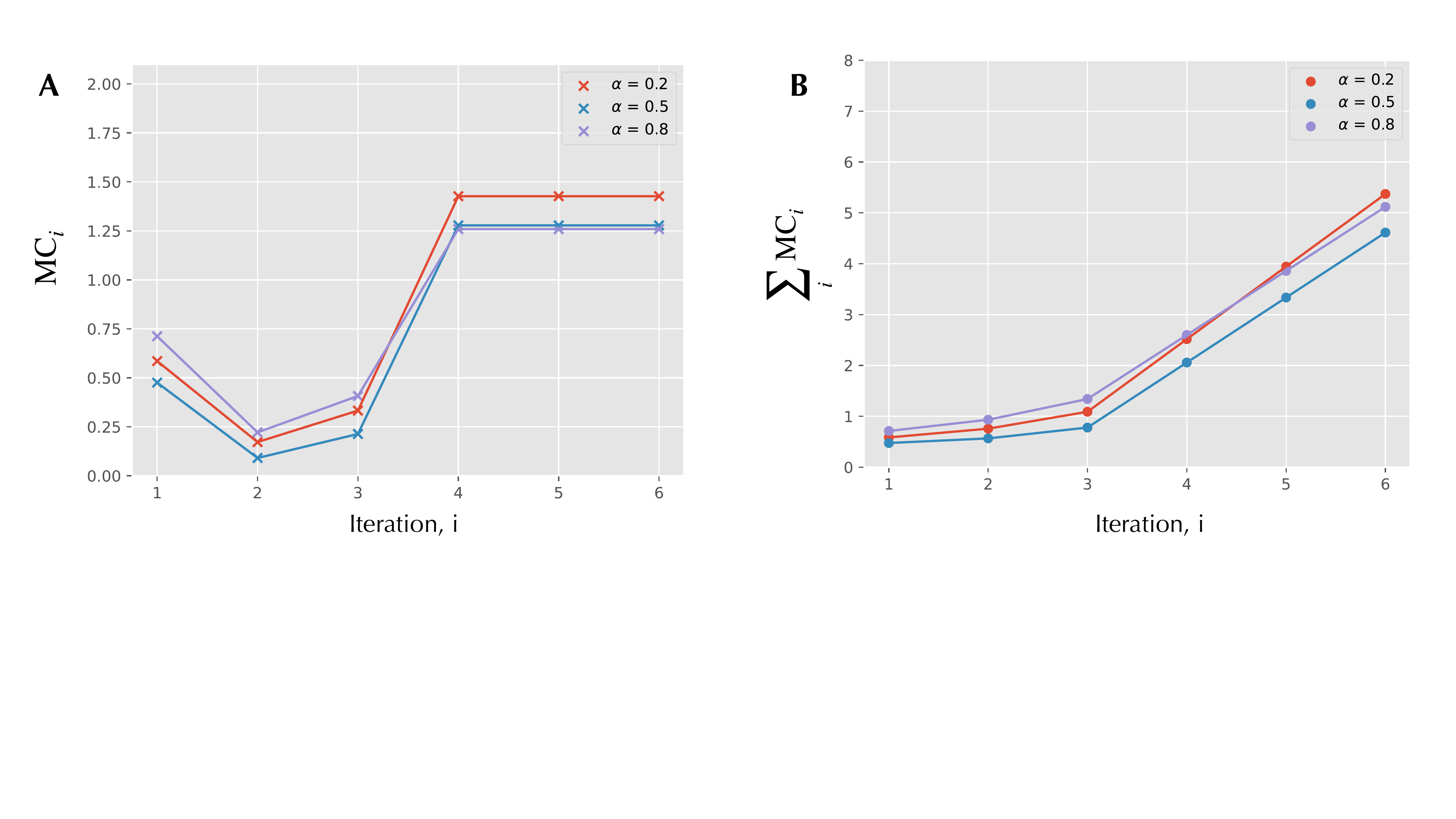}
    \caption{Mismatch cost of the \texttt{heaviside} program. The input variable \texttt{x} follows a binomial distribution $p_{\text{in}}(x = k) = \binom{n}{k} \alpha^k (1 - \alpha)^{n - k}$, parameterized by a Bernoulli parameter $\alpha$, while the variable \texttt{c} is fixed at 5. This input distribution induces a distribution over all program variables, thereby determining the initial distribution $p_{X_0}$ over the program's state space. A) Mismatch cost incurred in each iteration of the \texttt{heaviside} program. Regardless of the initial distribution, the program reaches a steady state within 4 iterations, after which the per-step MMC becomes constant. B) Cumulative mismatch cost over successive iterations of the \texttt{heaviside} program.}
    \label{fig_heaviside_mmc}
\end{figure*}

To model the computation and execution mechanism of stored-program computers, we use a simplified abstraction of stored program architecture known as Random Access Stored  Program (RASP) machines. Much like stored program architecture used in real digital computers, a RASP machine consists of a memory that stores both the program (the list of instructions) and the data the program operates on. {\color{black} It is important to clarify that the term ``random'' in RASP refers to random-access memory---i.e., the ability to access any register directly---and does not imply stochastic or probabilistic computation.}

We provide an example in Fig.~\ref{fig2}, where two high-level programs are translated into their corresponding low-level RASP representations, which closely resembles how an actual computer would execute them. In this representation, the use of registers to store variable values becomes explicit, and the program counter---responsible for tracking the control flow---is clearly visible. Each variable in the program maps to a register in the RASP, including a dedicated register for the program counter. The state of the program is defined by the values of all the RASP registers storing the data and the value of the program counter. For examples, the instantaneous state of program (a) in Fig.~\ref{fig2} is specified by the values of registers holding the variables \texttt{x, y, z}, and the program counter \texttt{pc}. 

Starting from initial input values, execution proceeds step by step by updating the contents of these registers. At each step, the program counter selects the next instruction to execute, and that instruction deterministically updates the contents of the registers and advances the counter. By repeating this clocked process, the machine follows a well-defined sequence of states corresponding to the execution of the program. 

As the program execution proceeds, the joint state of all the registers evolves through a sequence of configurations---effectively tracing a trajectory through the program’s state space. We can track the discrete sequence of values all the registers go through as the program runs on a given input. The role of the program counter is crucial in the description of the state of the program. At any moment during execution, the values of all variables alone do not fully specify the program’s behavior---one must also know which instruction is being executed. The same variable values can be modified differently depending on the program counter. 

This perspective allows us to model the state of a program as a node in a graph, and execution of instructions as directed edges on those nodes---each directed edge from one node to another is a transition associated with the computational step. By simulating the associated RASP of a program for every possible values of input variables, one can generate the adjacency matrix $G$ associated with the computational graph of the program. For example, the directed graph illustrated in Fig.~\ref{fig_heaviside_statespace} corresponds to the \texttt{heaviside} program described in~\ref{fig2}(b).

More formally, let $\X_\A$ denote the set of all valid joint states of the variables in algorithm $\A$ including the program counter; this set corresponds to the nodes of the computational graph. The variables in a program can generally be grouped into two categories: input variables and non-input (internal) variables. Within non-input variables, some variables such as loop counters, flags and program counter are special variables since they are initialized to fixed values (e.g., the program counter begins at 0). 

Let $\x_{in}$, $\x_{sp}$, and $\x_{nin}$ denote the joint state of input variables, special variables, and non-input variables that are not special, respectively. Let $X_0$ be the random variable representing the initial state of the program and let $X_i$ be the random variable representing the program state  after $i^{th}$-iteration of the associated RASP machine. Specifically, let $X_{n}$ be the random variable representing the program state after the program halts. We assume that the input variables are freshly sampled from $p_{\IN}$ at the start of each run. Special variables--such as flags, loop counters, and the program counter--are initialized to fixed values. All the other internal (non-input) variables, $\x_{nin}$, are assumed to retain the values they had at the end of the previous execution. Let $p_{\nin}$ be the marginal distribution over these non-input variables from the previous run, i.e,. $p_{\nin}(\x_{nin}) = \sum_{\x_{\IN}, \x_{\mathrm{sp}}}p_{X_n}(\x)$, where $\x$ is the joint state of {\it all} variables. Thus, the initial joint distribution over the full program state is:
\begin{equation}
    p_{X_0}(\x) = p_{\IN}(\x_{in}) \, \delta_{\mathrm{sp}}(\x_{sp}) \, p_{\nin}(\x_{nin}),\label{eq_initial_dist}
\end{equation}
where $\delta_{X_{sp}}(x_{sp})$ is the Kronecker delta, ensuring that the special variables are set to their predefined initial states with probability 1. 

Starting with $p_{X_0}$, the dynamics induced at the ensemble level from the state transitions of the RASP machine can be modeled naturally by the adjacency matrix,
\begin{equation}\label{eq_G_evolution}
    p_{X_{i+1}}(\x) = \sum_{\x' \in \X}G(\x|\x')p_{X_{i}}(\x'),
\end{equation}
or $p_{X_{i+1}} = G p_{X_i}$ as a short-hand. Therefore, the distribution over the state of the program after $i^{th}$ iteration is given by,
\begin{equation}
    p_{X_{i}} = G^i p_{X_0}.
\end{equation}
Thus, starting with an initial distribution $p_{X_0}$, the above equation provides the sequence of probability distributions over program's state space across each iteration. 

Note that for programs that do not have any conditionals, the number of steps required to finish the program does not change with the input. In that case, the $i^{th}$ iteration of the map $G$ corresponds to the state transition associated with the $i^{th}$ computational step of the program. The number of iterations of $G$ required for $p_{X_i}$ to reach a steady state is same as the number of computational steps needed to halt the program. However, programs that do involve conditionals, such as the \texttt{heaviside}, the number of steps to halt the program may depend on the input. {\color{black}In case of \texttt{heaviside}, this number is 6 if $x > c$ and 7 otherwise.} As a result, in general the iteration index $i$ of $G$ does not directly correspond to a specific computational step. Moreover, the number of iterations needed for $p_{X_i}$ to reach a steady state is given by the maximum execution length across all possible inputs. 
Nonetheless, these changes of states are driven by the same map $G$, which corresponds to repeating the same process across all steps of a program in a computer. 

\section{Mismatch cost of computer programs}\label{sec_MMCprograms}

Consider an adjacency matrix or map $G$ representing to the state space behavior of the RASP associated with a program. As described in previous section, changes of states are driven by the repeated application $G$, which corresponds to repeating the same process across all steps of a program in a computer. Let $q_X$ denote the prior distribution associated with the cost function of the map $G$. As discussed earlier, this prior remains fixed across iterations. Consequently, the MMC incurred during the $i^{\text{th}}$ iteration of the map $G$ is given by,
\begin{equation}
    \MMC_i(p_{X_0}) = D(p_{X_{i-1}}\|q_{X}) - D(G p_{X_{i-1}}\|Gq_{X}).\label{eq_iteration_MMC}
\end{equation}
The total MMC of running the entire program is given by the sum over all iterations until the program halts,
\begin{align}
    \MMC(p_{X_0}) &= \sum_{i =1}^n\MMC_i(p_{X_0}) \\
    &= \sum_{i = 1}^{n} \left[D(G^{i-1}p_{X_0}\|q_{X}) - D(G^{i}p_{X_0}\|Gq_{X})\right] \label{eq_sum_MMC}
\end{align}
To determine the MMC for a given computer program using Eq.~(\ref{eq_sum_MMC}), we can either start with illustrative choices of the function $f$ and derive the corresponding prior distribution using the methods outlined in App.~\ref{app_1}, or we can make illustrative assumptions about the prior directly. In the example of Heaviside program (Sec.~\ref{subsec_heaviside}) that follow, we adopt the first approach: we assume a uniform $f$ and compute the associated prior distribution from it. 

\subsubsection{Heaviside program}\label{subsec_heaviside}

Consider the \texttt{heaviside} program introduced in Fig.~\ref{fig2}, whose state space is depicted in Fig.~\ref{fig_heaviside_statespace}, generated through simulation of the corresponding low-level RASP code. The state of the program is given by the joint value of variables \texttt{x, c, a, pc}, where \texttt{x} and \texttt{c} are the input variables,  \texttt{a} is the conditional flag variable, and \texttt{pc} is the program counter. The program's state transitions are captured by the adjacency matrix of its state-space graph, as formalized in Eq.~(\ref{eq_G_evolution}). Given an initial distribution $p_{X_0}$, Eq.~(\ref{eq_G_evolution}) describes the discrete-time evolution of the distribution under repeated application of a fixed stochastic map $G$. 

We assume that the input variable \texttt{x} takes values in the set $\{0, \dots, 9\}$ and follows a binomial distribution $
p_{\text{in}}(x = k) = \binom{n}{k} \alpha^k (1 - \alpha)^{n - k},
$ parameterized by the Bernoulli parameter $\alpha$. For illustrative purposes, the variable \texttt{c} is fixed at 5. This input distribution induces a distribution over the program’s other variables, which in turn defines the initial state distribution $p_{X_0}$ over the program’s full state space, as described in Eq.~(\ref{eq_initial_dist}). By simulating the state transitions of associated RASP for every possible input values, we obtain $G$. The prior $q_X$ is obtained by assuming uniform $f$ over all states.

We use Eq.~(\ref{eq_iteration_MMC}) to compute the MMC incurred at each iteration, and Eq.~(\ref{eq_sum_MMC}) to compute the total MMC. Figure~\ref{fig_heaviside_mmc} shows both the stepwise and cumulative MMC results. As seen in panel (a), the program reaches a steady state after four iterations, beyond which the per-step MMC remains constant. As a result, panel (b) shows that the cumulative MMC grows linearly after the steady state is reached.

\begin{figure*}
    \centering
    \includegraphics[trim = {0 4cm 0 0}, width=0.8\linewidth]{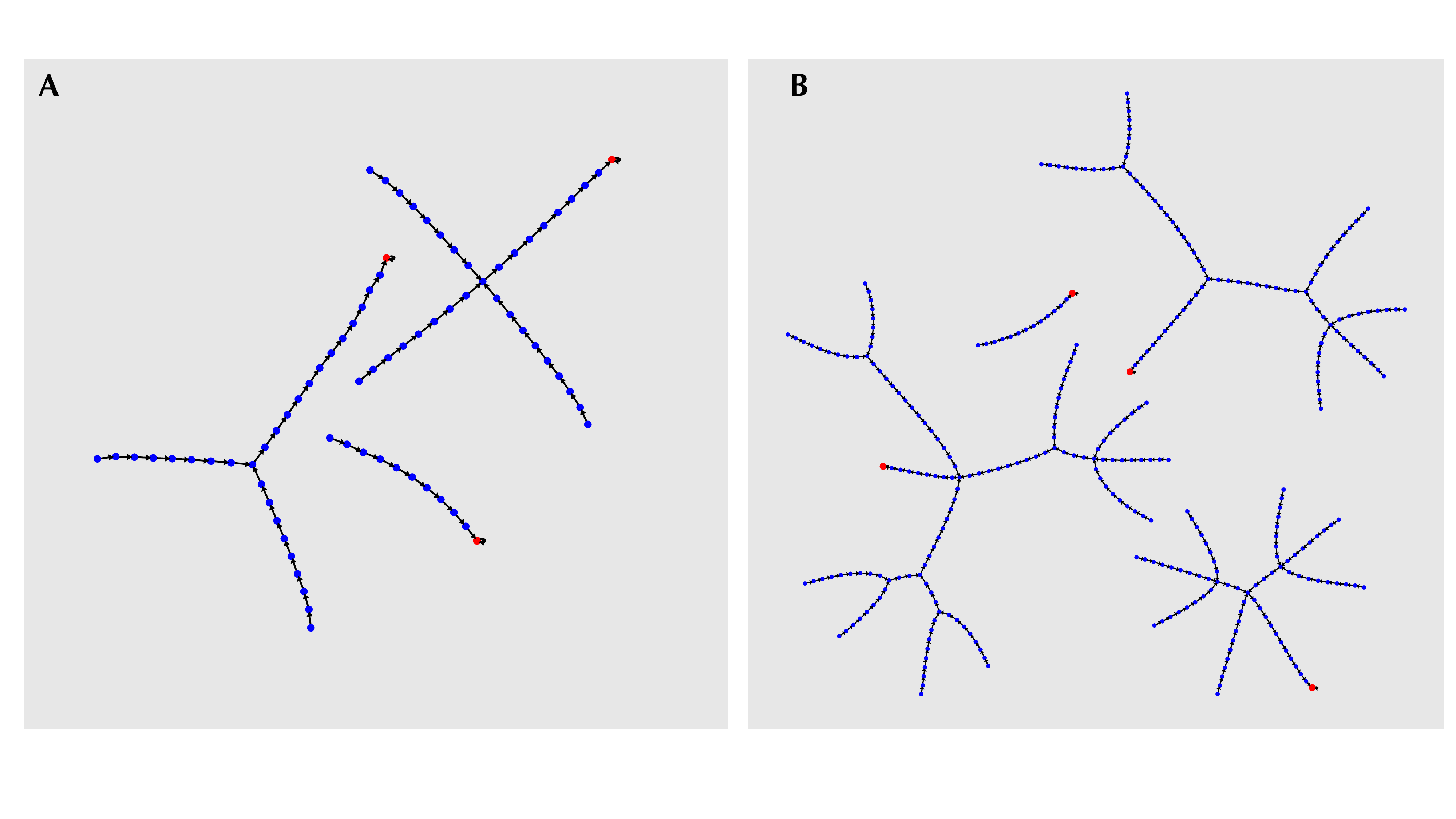}
    \caption{State space of the \texttt{BubbleSort} program for input arrays of length $n = 3$ (A) and $n = 4$ (B). The input arrays are permutations of $\{1, \dots, n\}$. Each node represents a unique state of the program, and directed edges correspond to state transitions caused by the execution of individual instructions. Leaf nodes represent possible initial states of the program, determined by different input permutations. As the program runs, it follows a deterministic trajectory through the state space, eventually reaching a terminal (halt) or attractor state for each initial condition. }
    \label{fig:dyn34}
\end{figure*}

\begin{figure}
    \centering
\includegraphics[width=0.9\linewidth]{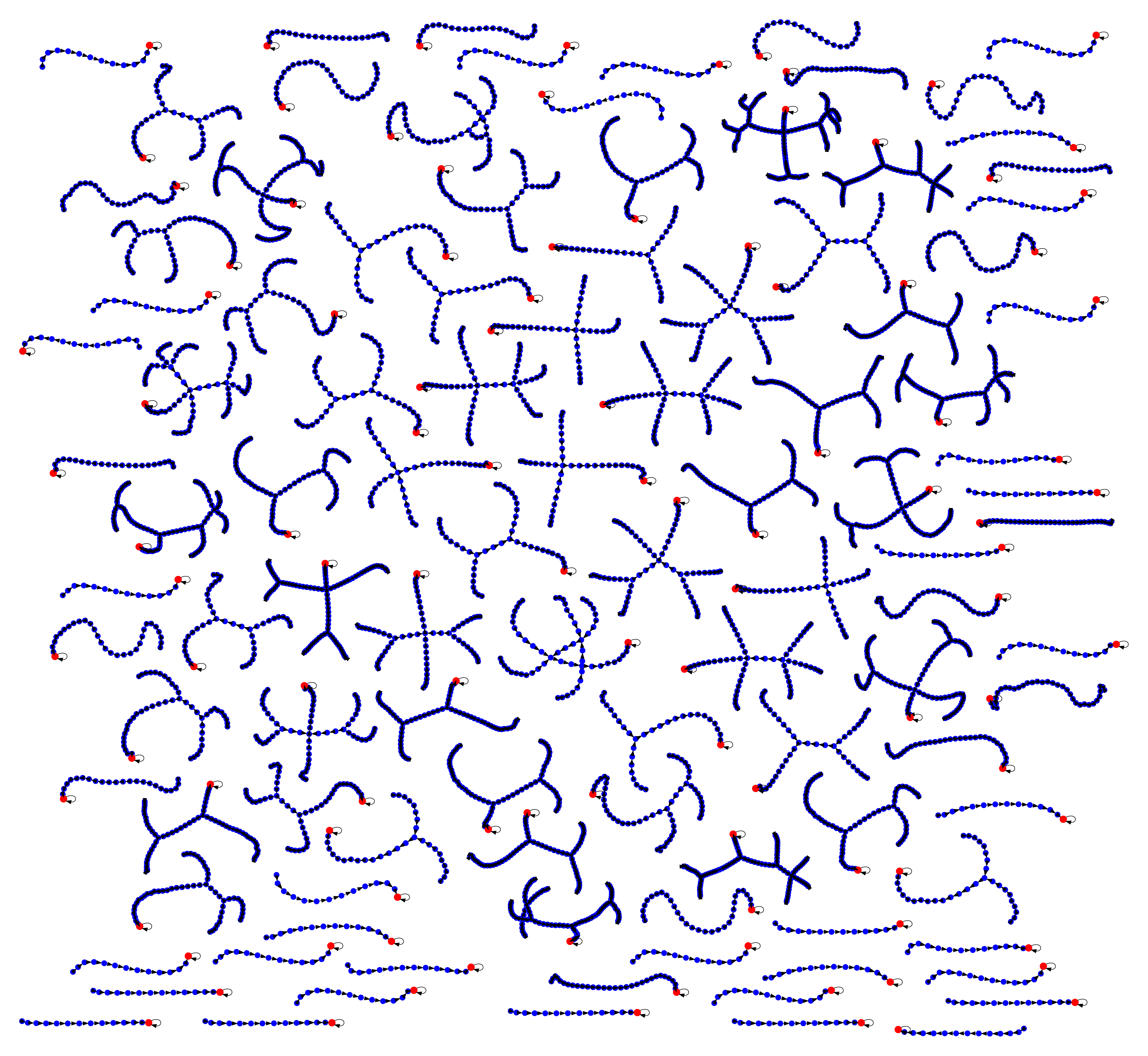}\\
\caption{State space of the \texttt{BubbleSort} program with input arrays allowing repeated entries from the set $\{1, \dots, n\}$. Compared to the case with distinct elements, the state space for $n = 4$ is significantly larger due to the increased number of possible input configurations, and the program contains many more fixed points (halt states).}
    \label{fig:combs4}
\end{figure}

\begin{figure*}
    \centering
    \includegraphics[width=0.98\linewidth, trim = {0 9cm 0 0}]{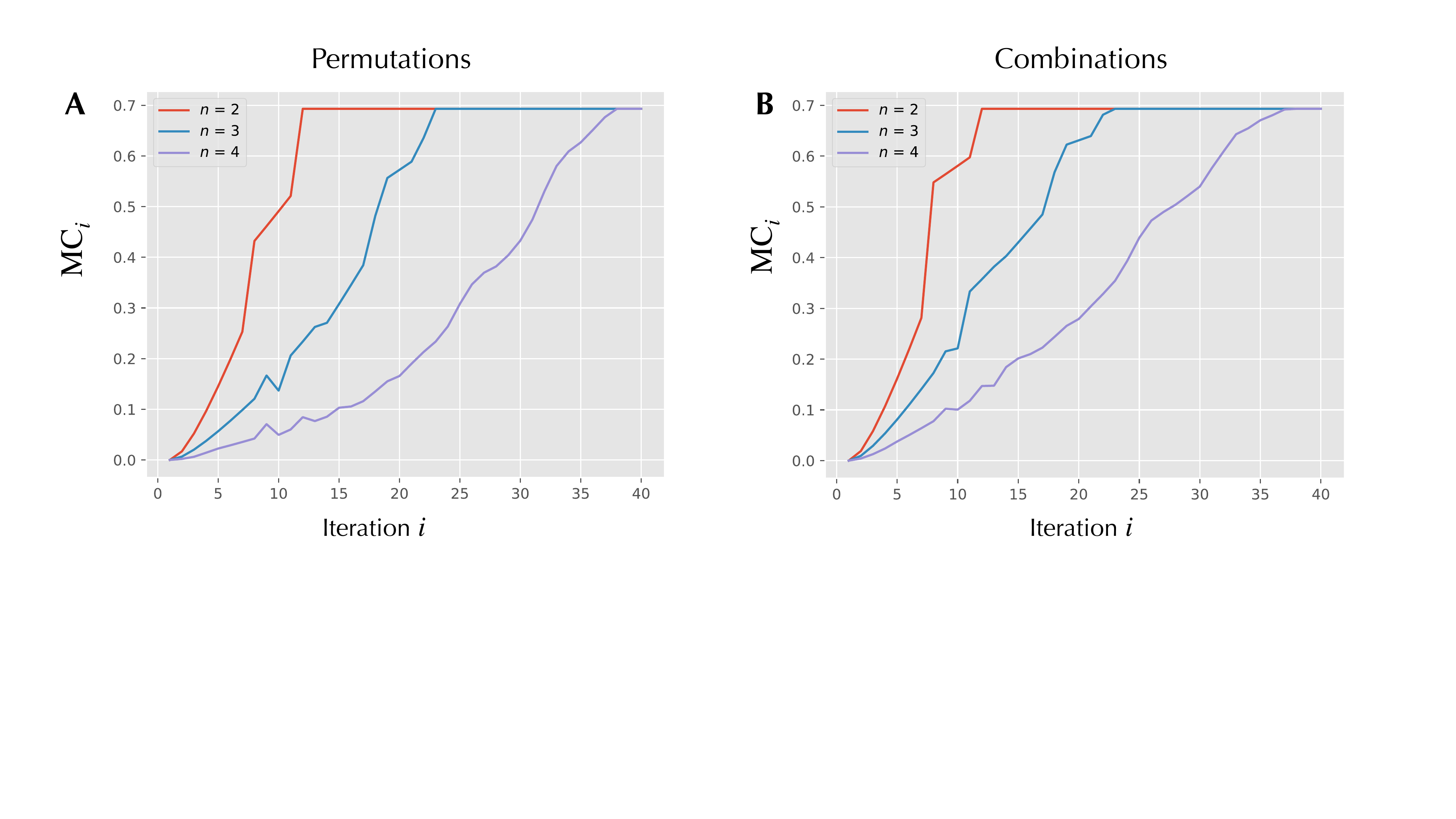}    
    \caption{ Mismatch cost incurred in each iterative step of the bubble sort algorithm, denoted by $\MMC_i$. Panel A shows the case where input arrays of length $n$ are all permutations of $\{1, \dots, n\}$, while panel B considers input arrays formed by combinations (with possible repetitions) of $\{1, \dots, n\}$. In both cases, the distribution over initial states is taken to be uniform, and the prior distribution $q_{X_0}$ is also assumed to be uniform over state space. }
    \label{fig:pmcpermcombs}
\end{figure*}

\subsubsection{Bubble sort program}\label{subsec:bubble}

We now turn to sorting programs for integer arrays of length $n$. Fig.~\ref{fig_bubble} in the Appendix depicts the source code for \texttt{BubbleSort} program and the corresponding RASP-like code. The state of the program consists of an array \texttt{arr} of length \texttt{n}, boolean flag \texttt{swapped}, and loop counters \texttt{i} and \texttt{j}. The full state of the program at any step is defined by the joint values of these registers along with the program counter. As instructions execute, both the register values and the program counter are updated, producing a sequence of state transitions.

Before proceeding to generate the state space of this program, we address two key considerations. First, for simple programs without arrays, the RASP model maps neatly onto low-level stored program architecture: each variable corresponds to a register, and the program’s state is defined by the joint state of all registers and the program counter. However, array-handling introduces complexity. In real systems, array elements are typically stored in memory and accessed via register-stored addresses. To avoid this complication, we adopt a simplified RASP-like model in which all program variables—including individual array elements—are treated as if directly stored in registers. This keeps our definition of the program state minimal and unified: a joint configuration of all variables and the program counter.

Second, the size of the joint state space grows rapidly with array size $n$. The dominant factor is the exponential number of possible input arrays. For instance, if array elements are digits in $\{0, 1, ..., 9\}$, the number of possible input arrays is $10^n$, leading to a total state space of size roughly $\sim n \cdot 10^n$ ($n$ accounting for internal variables). If the input array is restricted to be a permutation of $\{1, ..., n\}$, the number of inputs becomes $n!$, and the total state space scales as $\sim n \cdot n!$. In either case, the combinatorial explosion imposes practical limits on simulation of lower level RASP.

We immediately observe that this approach suffers from a combinatorial sampling challenge as the input array length increases. To manage this, we restrict our analysis to small input sizes, specifically $n = 2, 3, 4, 5$, and separately study the cases where the input array is a permutation of $\{1, \dots, n\}$ without repeated entries, and where it is a combination of it with repeated entries are allowed.

To construct the associated stochastic map $G$ of the bubble sort algorithm with restricted input arrays that are permutations of $\{1, \dots,n\}$, we simulate the program code for each input array and record the transitions in the program’s state after each instruction. We perform this simulation for arrays of length $n = 3$ and $n = 4$. The resulting phase spaces with state transitions are shown in Fig.~\ref{fig:dyn34}. The leaf nodes in these graphs represent all possible input arrays---$3! = 6$ for $n = 3$ and $4! = 24$ for $n = 4$.

For arrays of length $n$ with repeated entries drawn from the set $\{1, \dots, n\}$, the state space expands significantly since it includes input combinations with duplicates (e.g., $[1,1,3,3]$). The enlarged state space for $n = 4$ is shown in Fig.~\ref{fig:combs4}.

To evaluate the MMC, we assume a uniform prior in both settings: one over the state space of arrays with non-repeated entries (permutations) and one over arrays with repeated entries (combinations). We consider array lengths $n=2,3,4$ in each case. We also assume that the acutal initial distribution over the state space is uniform. The resulting MMC per iteration of the map $G$ is shown in Fig.~\ref{fig:pmcpermcombs}(a) for the permutation case and Fig.~\ref{fig:pmcpermcombs}(b) for the combination case. In both settings, the MMC per iteration approaches a steady value once the program halts, and the number of iterations required for halting increases with the input length.

\begin{figure}
    \centering
\includegraphics[width=0.9\linewidth]{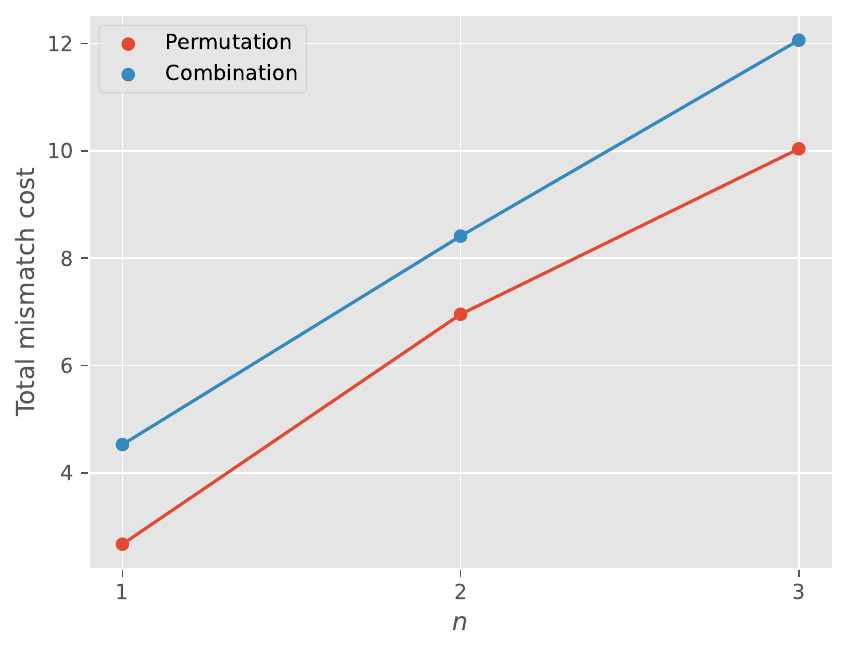}
    \caption{Aggregate MMC, obtained by summing over all iterations until termination $\left(\sum_{i = 1}^{\mathrm{halt}}\MMC_i\right)$ is plotted for bubble sort for both permutations (red) and combinations (blue) of inputs of length $n$. The initial distribution is taken to be uniform. The program has higher total MMC when the input arrays are allowed to have repetitions compared to the case where inputs are restricted to permutations.}
    \label{fig:pmccombsall}
\end{figure}

The aggregate MMC—obtained by summing over all iterations until termination—is plotted as a function of input size in Fig.~\ref{fig:pmccombsall}. The total MMC is consistently larger in the combination case (allowing repeated entries) than in the permutation case (distinct entries only). Allowing repeated entries affects the state space in two competing ways. On the one hand, repeated values can reduce the number of swaps: when adjacent elements are equal, no exchange occurs, allowing the algorithm to terminate in fewer steps on average, thereby reducing the number of transitions per run. On the other hand, allowing repetition substantially increases the number of possible input arrays, enlarging the overall state space. The net effect is a trade-off: repeated entries can shorten individual execution paths, yet increase the average MMC due to the larger and more complex state space over which the dynamics unfold.

% Figure~\ref{fig:pmcpermcombs} shows a comparison in the behavior of $\MMC_i$ for uniform input distributions $p_{X_0}$. 
% The value at which $\MMC_i$ becomes constant is the value at which the distribution $p_{X_0}$ has converged to a steady state value. In principle, comparisons should be done at that value. $\MMC_i$ increases monotonically because $G^k p_{X_0}\rightarrow p^{eq} \neq q_{X_0}$, e.g. it concentrates away from $q_{X_0}$ in $KL$-divergence terms. 

%We can also see the difference between inputs in the rates of EP mismatch production between permutations and combinations and for different values of $n$, the input vector size. What transpires is that PMC is a complicated measure to analyze. 

%Fig.~\ref{fig:pmccombsall} shows the cumulative MMC for $n\in \{2, 3, 4, 5\}$ as a function of iterations for both permutation case and combination case. %for inputs that are permutations and combinations. 

In the next subsection, we discuss how this framework can be extended to programs that invoke other programs as subroutines. As an illustrative example, we present the Bucket Sort algorithm, which partitions an input array into buckets and calls the bubble sort program to sort each sub-array \mbox{individually}.

\subsection*{Subroutine calls}

Consider two programs $\A$ and $\B$. State space associated with each program consists of the joint state of all variables in the program and the state of program counter. Let $\mathcal{X_A}$ and $\mathcal{Y_B}$ denote the state space of the program $\A$ and $\B$ respectively. Based on the framework discussed earlier in Sec.~\ref{sec_RASP}, one can define the dynamics of the distribution over the program's state space:
\begin{equation}
    p_{X_t} = G_A p_{X_{t-1}}, \quad p_{Y_t} = G_B p_{Y_{t-1}}
\end{equation}
where $G_A$ and $G_B$ are stochastic maps corresponding to $\A$ and $\B$. 

In a subroutine call of $\B$ by $\A$, one of the instructions in program $\A$ calls $\B$ to update the variables in $\A$.  We specifically focus on the call-by-value case, where program $\A$ passes a copy of the values to subroutine $\B$. Any changes made within the subroutine do not affect the original variables in the calling program. The variables within a subroutine are defined locally in its own scope, and the main program passes only the values---not references---of variables to the subroutine. As a result, the subroutine does not operate directly on the variables defined in the calling program but instead works on local copies of their values.

This allows us to treat the calling program and the subroutine---and their respective state spaces---largely independently. Specifically, we can define separate prior distributions for the stochastic maps $G_\A$ and $G_\B$, corresponding to the state spaces $\X_\A$ and $\X_\B$, respectively. Once the subroutine $\B$ is invoked within $\A$, and the input distribution to $\B$ induces an initial distribution $p_{Y_0}$ over its state space, the evolution of $p_{Y_0}$ proceeds independently of the state distribution of $\A$.

\begin{equation}
    p_{Y_t} = G_B p_{Y_{t-1}}
\end{equation}
This independence arises because, during the execution of program $\B$, all variables associated with program $\A$ remain unchanged. On the other hand, if we focus solely on the discrete-time updates of program $\A$'s variables---ignoring the internal dynamics of $\B$ during its invocation---the evolution of the distribution over $\A$'s state remains well-defined:

\begin{equation}
    p_{X_{t}} = G_A p_{X_{t-1}}
\end{equation}
Denoting the mismatch cost associated with a complete run (i.e., until the initial distribution reaches a steady state) of each program as $\MMC_A(p_{X_0})$ and $\MMC_B(p_{Y_0})$, we define the total mismatch cost of the joint program—where $\A$ calls $\B$—as the sum of the two:
\begin{equation}\label{eq_subroutine1}
    \MMC_A(p_{X_0}) + r \cdot \MMC_B(p_{Y_0})
\end{equation}
assuming that program $\B$ is called $r$ times within program $\A$, and that each call induces an initial distribution $p_{Y_0}$ over $\B$'s. 

A few important conditions must be satisfied for expression~(\ref{eq_subroutine1}) to serve as a valid MMC lower bound on the EP associated with the joint program execution. First, the initial distribution induced on the state space of $\B$ must be the same for each invocation. Second, the number of times the subroutine $\B$ is called must be independent of the input to program $\A$. Finally, the timing (i.e., the steps at which $\B$ is invoked within $\A$) must also be independent of the input to $\A$. Moreover, it is important to emphasize that the MMC in Eq.~(\ref{eq_subroutine1}) does not account for the thermodynamic cost associated with the creation and destruction of correlations between the variables of program $\mathcal{A}$ and subroutine $\mathcal{B}$ each time $\mathcal{A}$ calls $\mathcal{B}$ with new input values. Nonetheless, Eq.~(\ref{eq_subroutine1}) provides a valid lower bound on the cost function. 

We consider a simplified example of the Bucket Sort algorithm, where an input array is divided into two sub-arrays (buckets), and the bubble sort subroutine is called separately to sort each of these buckets.

\subsubsection*{Bucket-sort program}
Bucket sort is an efficient sorting algorithm that divides the input data into a fixed number of buckets, sorts the elements within each bucket---typically using another sorting algorithm or directly if the buckets are small---and then concatenates the sorted buckets to produce the final sorted output. Unlike bubble sort, which has a worst-case time complexity of $O(n^2)$, bucket sort can achieve an average-case time complexity of $O(n)$ under certain input distributions (e.g., when the input is uniformly distributed). 

\begin{figure}
    \centering
    \includegraphics[width=0.95\linewidth]{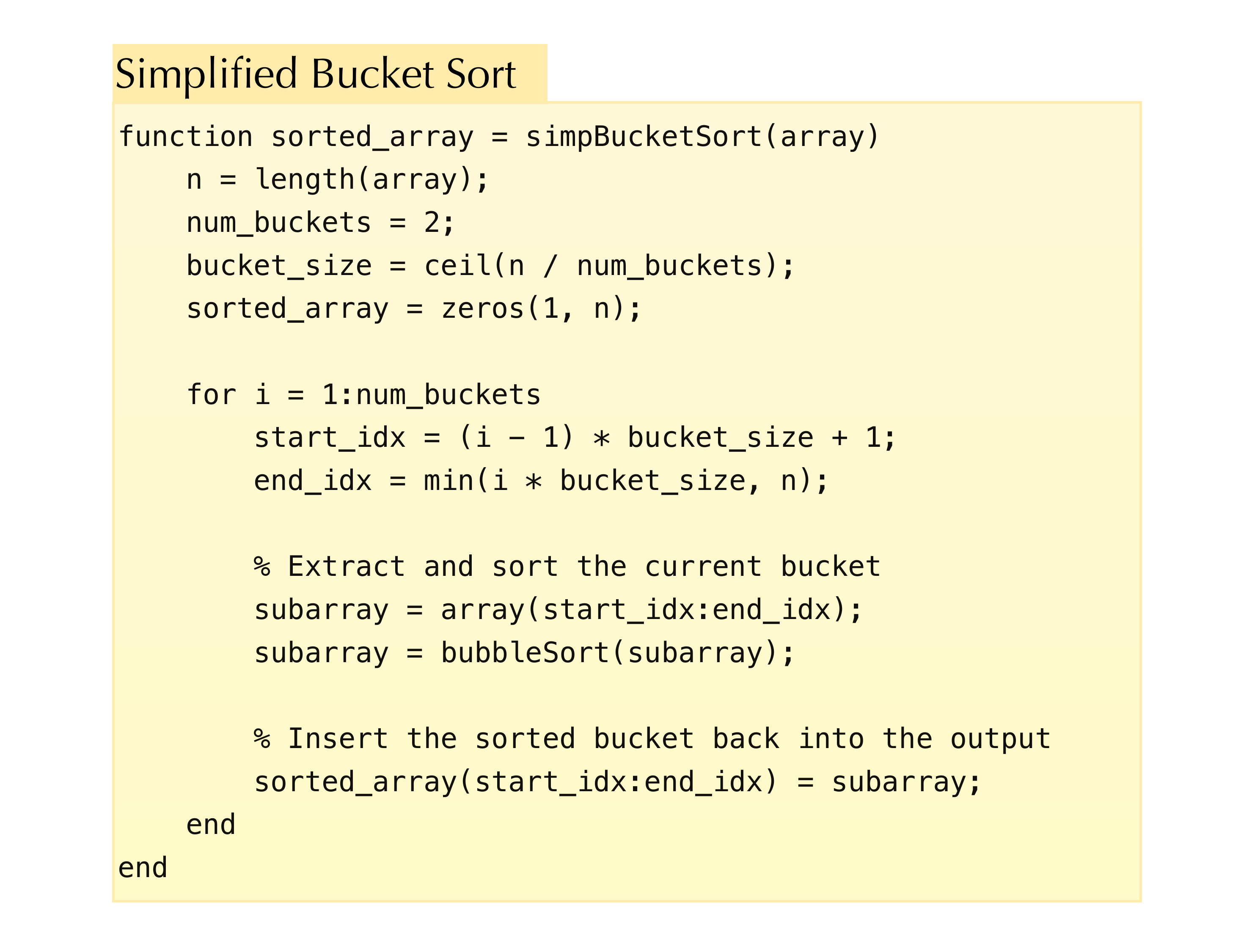}
    \caption{A source code for bucket sort written in MATLAB.}
    \label{fig_BucketSort_code}
\end{figure}

In this section, however, to illustrate the MMC of subroutine calls, we focus on a simplified version of bucket sort (shown in Fig.~\ref{fig_BucketSort_code}) that takes an input array of length $n$, divides it into two buckets (sub-arrays), calls \texttt{BubbleSort} to sort each sub-array, and then combines the sorted results. 

Note that the subroutine \texttt{BubbleSort} is called exactly twice, regardless of the input array. Furthermore, the timing of these calls is fixed and does not vary with the input. Moreover, the distribution over the variables within \texttt{BubbleSort} is the same for each call. Therefore, for the program in Fig.~\ref{fig_BucketSort_code}, we can apply Eq.~(\ref{eq_subroutine1}) to evaluate the MMC. It is also worth noting that Eq.~(\ref{eq_subroutine1}) does not account for correlations between the variables of the two subroutine calls. The MMC lower bound on the cost of the joint program defined in Fig~\ref{fig_BucketSort_code} is determined by
\begin{equation}
    \MMC_A(p_{X_0}) + 2\cdot \MMC_B(p_{Y_0})
\end{equation}
where $\A$ and $\B$ are the state spaces of \texttt{BucketSort} and \texttt{BubbleSort} respectively. Note that any initial distribution $p_{X_0}$ for the \texttt{BucketSort} program induces an initial distribution $p_{Y_0}$ for the \texttt{BubbleSort} program. 

\section{Discussion and Future work}\label{sec_conclusion}

{\color{black} The framework developed in this paper, which combines RASP machines with the MMC, is among the first attempts to address the thermodynamic cost of running computer programs in a unified manner without requiring detailed knowledge of the underlying device physics. It allows us to move toward a broader notion of algorithmic complexity that accounts not only for how long a program runs or how much memory it consumes, but also for the thermodynamic cost associated with its execution. 

For example, in the case of the \texttt{heaviside} program, we observe that the second and third iterations of the associated RASP machine incur substantially lower MMC than the first iteration (Fig.~\ref{fig_heaviside_mmc}(a)). This is not immediately obvious just by looking at the associated RASP alone (Fig.~\ref{fig2}(b)). However, examining the corresponding state space in Fig.~\ref{fig_heaviside_statespace}, the reason becomes intuitive: at the ensemble level, the first iteration results in a convergence of many distinct initial states into relatively few states. This contraction in state space at the distribution level leads to a larger MMC. A similar effect is visible in the fourth iteration. While this is only an intuitive explanation --- since the precise relationship between state-space convergence and MMC also depends on the prior distribution --- it suggests a broader principle: algorithmic steps that induce greater state-space convergence tend to incur higher MMC.

The \texttt{BubbleSort} examples illustrate an even more nuanced interaction between input structure, state space, and MMC. One might reasonably expect that \texttt{BubbleSort} acting on an array of size 2 would incur less MMC in each iteration than the same algorithm acting on arrays of size 3 or greater. Surprisingly, Fig.~\ref{fig:pmcpermcombs} shows the opposite: the MMC for $n = 2$ is consistently larger for at each iteration before the program halts compared to the MMC per iteration for $n = 3,4,5$. This cannot be explained by state-space structure alone. However, the number of iterations it takes the program to halt increases with increasing length of input array, and consequently the aggregate MMC is larger for larger  Since we assumed a uniform prior distribution, both the state space encoded in the stochastic map $G$ and the choice of $f$ together determine the prior, and hence the MMC. Isolating the precise cause of this behavior---and more generally, understanding how algorithmic structure, input size, and prior interact to determine MMC---remains an interesting direction for future work.
}
%{\color{purple} Refree: I would like to see a discussion of what the results on the bubble sort and bucket sort programs actually teach us. For instance: why does the steady MMC depend on the array size in Fig. 6? Are loops especially costly? Is there any relationship between MMC and logical irreversibility? Could parallel computing reduce the MMC?}

%The framework we established naturally extends to more complex algorithms involving loops, recursion, and intricate branching structures. This opens avenues to study more advanced computer programs such as dynamic programming, graph algorithms, or even machine learning pipelines. In all these cases, explicitly tracking the program counter remains essential for connecting algorithmic execution with transitions within the underlying program state space.

{\color{black}
One aspect we did not investigate in this paper is the relationship between MMC and other standard complexity measures such as time and space complexity. For instance, as discussed earlier, programs or algorithms that involve conditionals or loops generally have halting times that depend on the input: some inputs cause them to run longer than others. Given a probability distribution over inputs, one can therefore speak of the expected running time of a program before it halts. Every probability distribution over inputs is associated with such an expected time. A natural question for future work is whether the total MMC associated with an input distribution is related in any systematic way to the corresponding expected running time induced by the same prior distribution.

Because the methods developed here are, in many respects, first of their kind, they also come with shortcomings that we highlight and that future work may hope to address. The first, and more pressing, difficulty is the rapid growth of the state space as the number of program variables increases. This combinatorial explosion limits the practical applicability of the current method to relatively small programs. At present, the authors do not know a general way around this obstacle. However, one promising direction would be to replace exhaustive enumeration of the entire state space with probabilistic sampling techniques---such as Monte Carlo methods---which approximate key quantities by sampling a manageable subset of states rather than computing over all of them.

The second issue concerns the estimation of the prior distribution, which plays a central role in the calculation of MMC. As discussed in the paper, one can either begin with a reasonable guess for the function $f(x)$ and infer a prior from it, or directly attempt to estimate the prior distribution. In either case, the reliability of the final MMC value is only as strong as one’s confidence in the underlying estimation procedure. 
}

One promising direction for identifying prior-independent minimal costs involves analyzing the structure of the periodic mismatch cost. Consider Eq.~(\ref{eq_sum_MMC}), which has played a central role throughout this paper in defining the total mismatch cost for a given prior $q_X$:

\begin{equation}
 \MMC_{q_X}(p_{X_0}) = \sum_{i = 0}^{r-1} \left[D(G^{i-1}p_{X_0} \| q_X) - D(G^i p_{X_0} \| Gq_X)\right]
\end{equation}
As the system evolves through a sequence of state distributions $\{p_{X_0}, p_{X_1}, \dots, p_{X_r}\}$ according to the update rule $p_{X_{t+1}} = G p_{X_t}$, there exists a distribution $\hat{q}_X$ that minimizes the sum above. This optimal prior $\hat{q}_X$ defines a special mismatch cost $\MMC_{\hat{q}_X}(p_{X_0})$, which acts as a strictly positive lower bound on the MMC incurred for any other choice of prior $q_X$. Importantly, this strictly positive minimal MMC is completely independent of the underlying physical process and arises solely from the computational map $G$ and its repeated application. %While this insight points toward a promising path for establishing prior-independent minimal thermodynamic costs, we do not explore it further here and leave a rigorous treatment for future work.

% Moreover, given a set $\{p_i\}$ of possible initial distributions, one could seek to establish a universal bound relating the minimal and maximal mismatch costs across these distributions, independent of the choice of prior $q$. \AY{If $\MMC_q$(p) denote the MMC of a process with the prior distribution $q$ on an actual distribution $p$ ... } Specifically, one could aim to find bounds on the relationship between
% \begin{equation}
%     \min_{q} \MMC_q(p)
% \quad \text{and} \quad
% \max_{i,j} \big[\MMC(p_i \parallel q) - \MMC(p_j \parallel q)\big],
% \end{equation}
% which hold for any prior $q$. Such a bound would directly translate to constraints on the thermodynamic entropy production, relating
% \begin{equation}
%     \min_i \C(p_i)
%     \quad \text{and} \quad
%     \max_{i,j} \big[\C(p_i) - \C(p_j)\big].
% \end{equation}
% Establishing such universal relations could provide further insight into the thermodynamic costs across a range of initial conditions, regardless of detailed physical assumptions or priors.

Our current treatment of subroutine calls is restricted to scenarios where both the timing and frequency of the calls are fixed and independent of the input instance. However, in general programs, the stochasticity of input---when input is drawn from a distribution---can induce stochasticity in both when and how often a subroutine is invoked. That is, the call structure becomes input-dependent. Extending our current consideration to handle this more general case poses a very challenging and mathematically reach problem.

{\bf Acknowledgements.} This manuscript was posted on a preprint: \url{https://doi.org/10.48550/arXiv.2411.16088}. The work by FC was conducted under the auspices of the National Nuclear Security Administration of the United States Department of Energy at Los Alamos National Laboratory (LANL) under Contract No. DE-AC52-06NA25396.  FC was also financed via DOE LDRD grant 20240245ER. AY and DHW were supported by US NSF Grant CCF-2221345, and thanks the Santa Fe Institute for support.

{\bf Data Availability.}
All data, code, and analysis scripts used in this study are publicly available at \url{https://github.com/Kensho28/RASP}. No proprietary or confidential data were used.

\appendix
\onecolumngrid

\section{Methods of estimation of prior distribution}\label{app_1}

Consider a system with discrete space $\X$ and let $\simp$ denote the simplex on the state space. Suppose that an initial distribution $p_{X_0}\in \simp$ transforms to a final distribution $p_{X_1} \in \simp$, where $X_0$ and $X_1$ are random variables representing system's state at initial and final time respectively, such that the transformation can be expressed by a conditional distribution $G(x|x')$ specifying the probability of a final state given an inital state:
\begin{equation}
    p_{X_1}(x) = \sum_{x'\in\X} G(x|x')p_{X_0}(x')
\end{equation}
or $p_{X_1} = Gp_{X_0}$ as short hand. For a real valued function $f: \X \to \mathbb{R}$, consider a cost function of the form,
\begin{equation}\label{eq:b1}
\C(p_{X_0}) = S\left(Gp_{X_0}\right) - S\left(p_{X_0}\right)+ \avg{f}_{p_{X_0}} .
\end{equation}
Define an equivalence relation on the state space $\X$ as follows:

\begin{equation}\label{def_island}
x \sim x' \iff \exists y \in \X \text{ such that } G(y|x) > 0 \text{ and } G(y|x') > 0.
\end{equation}
Let $\mathcal{L}_G(\X)$ denote the partition of $\X$ induced by the transitive closure of this equivalence relation. This partition is referred to as the island decomposition of $\X$ under the map $G$, and each subset $\mathcal{Z} \in \mathcal{L}_G(\X)$ is called an island of $\X$ with respect to $G$. 

It is known that if $G$ induces a single island---i.e., $\mathcal{L}_G(\X) = \{\X\}$---then the minimizer of the cost function $\C$ in Eq.~\ref{eq:b1}, given by
$q_{X_0} = \mathrm{arg}\min_{r_0} \C(r_0)$, is unique~\cite{wolpert2020thermodynamics}.

However, when the map $G$ induces multiple islands---each indexed by $c$ and denoted by $\mathcal{Z}_c$---we can define, for any distribution $p_{X_0} \in \simp$, the conditional distribution over each island as follows:

\begin{equation}
p_{X_0|c}(x) =
\begin{cases}
    \dfrac{p_{X_0}(x)}{p(c)} & \text{if } x \in \mathcal{Z}_c, \\
    0 & \text{otherwise}.
\end{cases}
\end{equation}
where $p(c) = \sum_{x \in \mathcal{Z}_c} p_{X_0}(x)$ is the total probability mass assigned to island $\mathcal{Z}_c$. The corresponding cost function restricted to island $\mathcal{Z}_c$ is then given by:

\begin{equation}\label{def_island_cost}
\C(p_{X_0|c}) = S\left(G p_{X_0|c} \right) - S\left(p_{X_0|c} \right) + \langle f \rangle_{p_{X_0|c}}.
\end{equation}

\noindent
Furthermore, the total cost can be expressed as a weighted sum over the island-restricted costs:

\begin{equation}
\C(p_{X_0}) = \sum_{c} p(c) , \C(p_{X_0|c}),
\end{equation}

\noindent
It has been shown that each island-restricted cost function $\C(p_{X_0|c})$ admits a unique minimizer~\cite{wolpert2020thermodynamics}, denoted by

$$
q^c_{X_0} = \mathrm{arg}\min_{r_0 \,:\, \mathrm{supp}(r_0) \subseteq \mathcal{Z}_c} \C(r_0).
$$

\subsection{Iterative method}
\label{app_prior}
In order to find the minimum of $\C(p_{X_0})$ in Eq.~\ref{eq:b1} subject to the constrain that $\sum_{x\in X} p_{X_0}(x) = 1$, we use the method of Lagrange multipliers. Evaluating the partial derivative of $\C(p_{X_0})$ with respect to $p_{X_0}(x')$ for an $x' \in X$;
\begin{equation}
    \der \C(p_{X_0}) = f(x') - \sum_{x\in X} G(x|x') \ln Gp_{X_0}(x)   + \ln p_{X_0}(x')
\end{equation}
The normalization constraint $g(p_{X_0}) = \sum_x p_{X_0}(x) - 1 = 0$ introduces a Lagrange multiplier $\lambda$, and the prior distribution satisfies the following equation:
\begin{equation}
    \der \C(q_{X_0})\bigg\rvert_{q_{X_0}} + \lambda \der g(p_{X_0})\bigg\rvert_{q_{X_0}} = 0
\end{equation}
This provides us an implicit equation for $q_{X_0}(x)$,
\begin{equation}\label{eq_lagrange2}
    \ln q_{X_0}(x) + f(x) - \sum_{x'\in X} G(x'|x) \ln Gq_{X_0}(x') + \lambda = 0,
\end{equation}
which can also written as,
\begin{equation}\label{eq:32}
    q_{X_0}(x) = e^ {-\lambda - f(x)} \left[\prod_{x'\in \X} (Gq_{X_0}(x'))^{G(x'|x)}\right].
\end{equation}
This equation is analytically solvable for only a few simple cases such as when $G$ is a permutation or when $G$ is complete erasure. However, for any $G$ and $f(x)$, the solution of Eq.~\ref{eq:32} can be found numerically. Consider the iterative version of Eq.~\ref{eq:32},
\begin{equation}\label{eq:33}
    q^{n+1}_{X_0}(x) = e^ {-\lambda^n - f(x)} \left[\prod_{x'\in \X} (Gq^n_{X_0}(x'))^{G(x'|x)}\right],
\end{equation}
where $\lambda^n$ ensuring that $q^{n+1}_{X_0}$ is normalized. The fixed point of Eq.~\ref{eq:33} is unique when the map $G$ has a single. It can be found numerically by starting with a random distribution and iteratively applying the map in~\ref{eq:33} to generate a sequence of distribution. The sequence coverages towards to prior distribution. 

%Moreover, when the map $G$ has multiple islands\ccc.
\subsection{Monte Carlo method}
To minimize the cost function~\ref{eq:b1}, we employ a simulated annealing algorithm that iteratively refines a candidate distribution over the simplex $\simp$ while gradually reducing randomness to ensure convergence. Start with a randomly initialized probability vector $q \in \simp$. Set the initial temperature $T_0$, cooling rate $\alpha < 1$, and interpolation parameter $s_0 \approx 1$. For each step $ n =\{1, 2, \dots \}$:
\begin{enumerate}
    \item Sample a new random distribution $ r \in \simp $.
    \item Form a candidate distribution:
    \[
    q' = (1 - s)r + s q,
    \]
    and normalize if necessary.
    \item Evaluate the cost:
    \[
    \C(q') = S(Gq') - S(q') + \langle f \rangle_{q'}.
    \]
    \item Accept $ q' $ with probability:
    \[
    P = 
    \begin{cases}
        1, & \text{if } \C(q') < \C(q), \\
        \exp\left(-\frac{\C(q') - \C(q)}{T}\right), & \text{otherwise}.
    \end{cases}
    \]
    \item If accepted, set $ q \leftarrow q' $ and update the best cost.
    \item Update the temperature: $ T \leftarrow \alpha T $, and reduce $ s $ gradually.
\end{enumerate}

The algorithm terminates after a fixed number of iterations or when successive updates no longer improve the cost. The final distribution $ q $ approximates the minimizer of $ \C(p_{X_0}) $. This annealing strategy ensures global exploration early on and local refinement as the temperature cools, making it suitable for non-convex optimization over the simplex.
The implementation of both methods for optimizing the cost function, including simulated annealing and the fixed-point iteration, can be found at the following repository: \url{https://github.com/Kensho28/RASP/blob/main/RASPdag/prior_finding%20(1).ipynb}.

\begin{figure*}
    \centering    \includegraphics[width=1\linewidth]{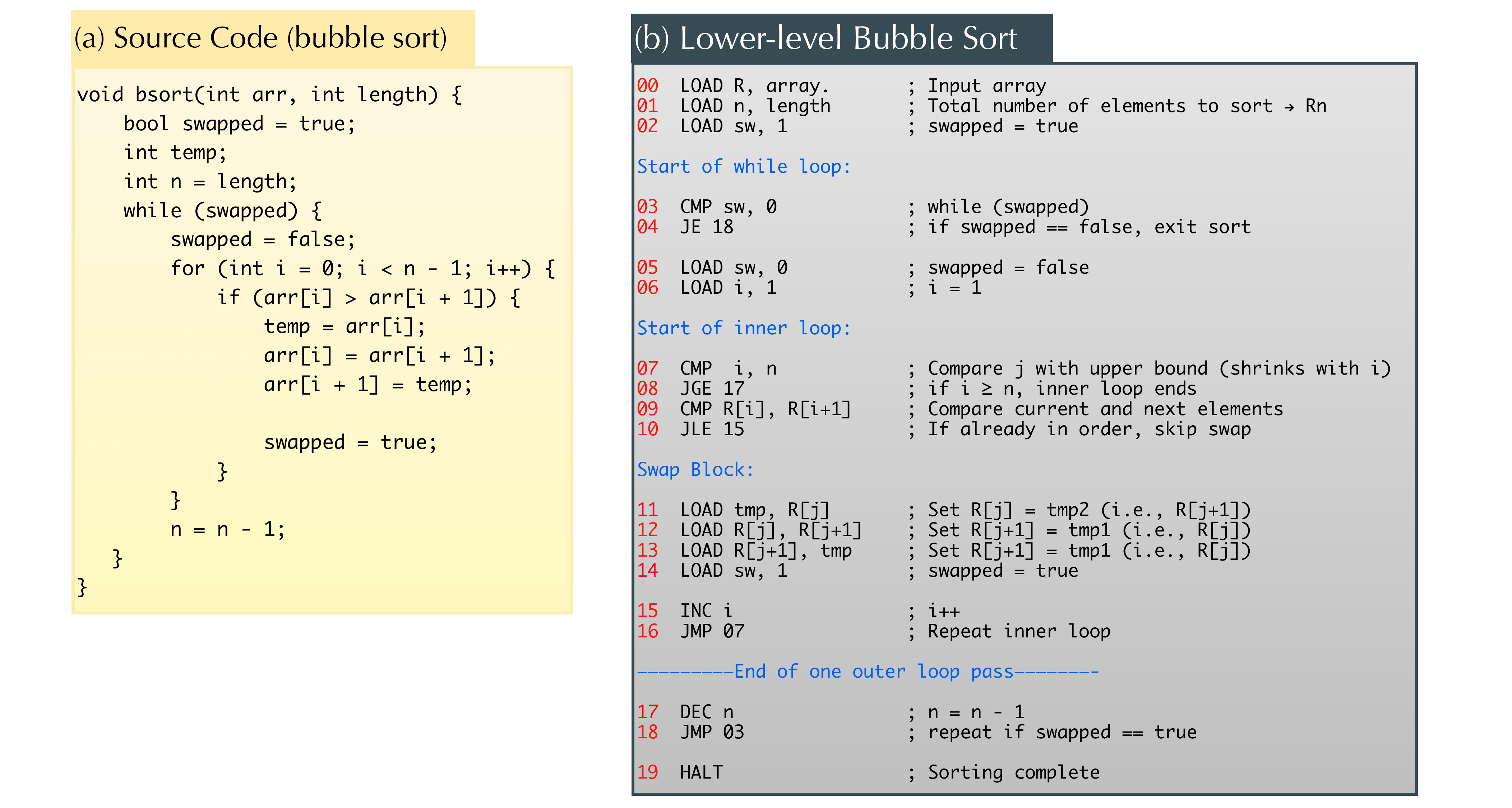}
    \caption{(a) High-level source code for the bubble sort algorithm written in C.
    (b) Corresponding lower-level representation of the bubble sort program, showing how the values of all variables and the program counter evolve during execution. The input array is stored in register \texttt{R}, and its length in \texttt{n}. Registers \texttt{sw}, \texttt{i}, and \texttt{tmp} serve as scratch registers corresponding to the boolean flag \texttt{swapped}, the loop counter \texttt{i}, and the temporary variable \texttt{temp}, respectively. The full state of the program at any step is defined by the joint values of these registers along with the program counter. As instructions execute, both the register values and the program counter are updated, producing a sequence of state transitions.}
    \label{fig_bubble}
\end{figure*}

\begin{table*}
\begin{center}
\begin{tabular}{ |p{4cm}||p{13cm}|}
 \hline
 \multicolumn{2}{|c|}{Commands used in the RASP} \\
 \hline
 \centering Symbol & Definition \\
 \hline
    \centering \texttt{LOAD Rn, value}   & Load a constant or memory value into register \texttt{Rn} \\
    \hline 
    \centering \texttt{STORE Rn, addr}   & Store value in register \texttt{Rn} to memory address \texttt{addr} \\
    \hline 
    \centering \texttt{READ Rn, addr}    & Read memory value at \texttt{addr} into register \texttt{Rn} \\
    \hline  
    \centering \texttt{ADD R1, R2, R3}   & \texttt{R3} $\leftarrow$ \texttt{R1 + R2} \\
    \hline 
    \centering \texttt{SUB R1, R2, R3}   & \texttt{R3} $\leftarrow$ \texttt{R1 - R2} \\
    \hline 
    \centering \texttt{MUL R1, R2, R3}   & \texttt{R3} $\leftarrow$ \texttt{R1 $\times$ R2} \\
    \hline 
    \centering \texttt{CMP R1, R2}       & Compare \texttt{R1} and \texttt{R2}; set condition flags \\
    \hline 
    \centering \texttt{JMP addr}         & Unconditional jump to address  \texttt{addr} \\
    \hline 
    \centering \texttt{JGE addr}         & Jump to \texttt{addr} if previous comparison was $\geq$ \\
    \hline 
    \centering \texttt{JLE addr}         & Jump to \texttt{addr} if previous comparison was $\leq$ \\
    \hline 
    \centering \texttt{INC Rn}           & Increment \texttt{Rn} by 1 \\
    \hline 
    \centering \texttt{DEC Rn}           & Decrement \texttt{Rn} by 1 \\
    \hline 
    \centering \texttt{CALL addr}        & Call subroutine at \texttt{addr}; push return address \\
    \hline 
    \centering \texttt{HALT}             & Stop execution \\
    \hline 
    \centering \texttt{R[n]}             & (Informal) Register array-like access notation \\
    \hline 
\end{tabular}
\end{center}
\caption{Table of notation RASP commands used in the paper.}\label{TABLE_RASP}
\end{table*}

\section{Proofs}

\subsection{Lower bound on the worst-case mismatch cost}\label{app:mismmin}

To derive a lower bound on the MMC as a function of the state-dependent term $f$, we focus on the case where the map $G$ induces a single island decomposition. In this setting, the prior distribution $q_{X_0}$ is unique.

The prior distribution $q_{X_0}$ satisfies Eq.~\ref{eq_lagrange2}, derived using the method of Lagrange multipliers:

\begin{equation}\label{eq_lagrange3}
\ln q_{X_0}(x) + f(x) - \sum_{x' \in \X} G(x'|x) \ln \left[G q_{X_0}(x')\right] + \lambda = 0,
\end{equation}
where $\lambda$ is the Lagrange multiplier associated with the normalization constraint on $q_{X_0}$. Once we re-write $g(x) = \sum_{x'\in X} G(x'|x) \ln Gq_{X_0}(x')$, the value of $\lambda$ is given by,
% \begin{equation}\label{eq_lagrange3}
% \ln{q_{X_0}(x)} + f(x) - g(x) + \lambda = 0.
% \end{equation}
% Using the constraint $ \sum_{x\in\X}q_{X_0}(x) = 1 $, we obtain:
\begin{equation}
\lambda = \ln\left( \sum_x \exp\left(g(x) - f(x)\right) \right).\label{eq:b4}
\end{equation}

Note that $-\lambda$ corresponds to the residual cost. To see this, multiply both sides of Eq.~\ref{eq_lagrange2} by $q_{X_0}(x)$ and sum over all $x \in \X$. This yields:
\begin{align}
    -\lambda &= \sum_{x\in\X}q_{X_0}(x)\ln{q_{X_0}(x)} + \sum_{x\in\X}q_{X_0}(x)f(x) - \sum_{x'\in X}  Gq_{X_0}(x') \ln Gq_{X_0}(x') \\
     &= - S(q_{X_0})+\avg{f}_{q_{X_0}} + S(Gq_{X_0})  \\
     &= \C(q_{X_0})
\end{align}
For any other initial distribution $p_{X_0} \in \simp$, the mismatch cost (MMC) is defined as the difference between the total cost incurred by $p_{X_0}$ and the minimal cost achieved by the prior distribution $q_{X_0}$:

\begin{equation}
\MMC(p_{X_0}) = \C(p_{X_0}) - \C(q_{X_0}).
\end{equation}

Since the cost function $\C(p_{X_0})$ is convex over the probability simplex $\simp$, its maximum is attained at one of the vertices---that is, at a distribution $\delta_c \in \simp$ of the form:

\begin{equation}
\delta_c(x) =
\begin{cases}
1 & \text{if } x = c, \\
0 & \text{otherwise}.
\end{cases}
\end{equation}
At these corners, the cost evaluates to:
\begin{equation}
\C(\delta_c) = f(c) - \sum_{x \in \X} G(x|c) \ln G(x|c).
\end{equation}
Since the term $-\sum_{x \in \X} G(x|c) \ln G(x|c)$ is upper bounded by $\ln |\X|$, it follows that if the function $f$ satisfies the condition

\begin{equation}\label{eq:b9}
\max_{x \in \X} f(x) - f(c) > \ln |\X|,
\end{equation}
for any $c$ such that $f(c) \neq \max_{x \in \X} f(x)$, then the maximum of the cost function over the corners is attained at the state where $f$ achieves its maximum:

\begin{equation}
\mathrm{arg}\max_c \C(\delta_c) = \mathrm{arg}\max_c f(c).
\end{equation}
This implies that the maximum of the cost function $\C$ is achieved at the vertex of the simplex corresponding to the state where $f(x)$ attains its maximum. Let us denote this state by
$$
u = \mathrm{arg}\max_{x \in \X} f(x).
$$
When condition~\ref{eq:b9} is satisfied, the maximum value of the cost function is given by:
\begin{equation}
\C(\delta_u) = f(u) - \sum_{x \in \X} G(x|u) \ln G(x|u), \label{eq:b11}
\end{equation}
Using Eqs.~\ref{eq:b4} and~\ref{eq:b11}, we can now derive a lower bound on the worst-case mismatch cost:
\begin{align}
\MMC^* &= \C(\delta_u) - \lambda, \label{eq:b13} \\
&= f(u) - \sum_{x \in \X} G(x|u) \ln G(x|u) + \ln \left( \sum_{x} \exp\left( g(x) - f(x) \right) \right),
\end{align}
where $g(x) := \sum_{y} G(y|x) \ln (G q_{X_0}(y))$. Applying the log-sum-exp inequality,
\begin{equation}
\ln \left( \sum_x \exp(g(x) - f(x)) \right) \ge \max_{x \in \X} \{ g(x) - f(x) \},
\end{equation}
and noting that
$$
\max_{x \in \X} \{ g(x) - f(x) \} = -\min_{x \in \X} \{ f(x) - g(x) \},
$$
Since $-g(x) = -\sum_y G(y|x) \ln (G q_{X_0}(y)) \le \ln |\X|$, when the condition~\ref{eq:b9} is satisfied, the minimum can be written explicitly as
\begin{equation}
\min_{x \in \X} \{ f(x) - g(x) \} = f(v) - g(v),
\end{equation}
where $v = \mathrm{arg}\min_{x \in \X} f(x)$. Substituting this expression back into Eq.~\ref{eq:b13} yields:
\begin{equation}
\MMC^* \ge f(u) - \sum_{x \in \X} G(x|u) \ln G(x|u) - f(v) + g(v).
\end{equation}
We can also write this in terms of the max–min gap in $f$, giving:
\begin{equation}\label{eq:B20}
\MMC^* \ge \max_{x} f(x) - \min_{x} f(x) + g(v) - \sum_{x \in \X} G(x|u) \ln G(x|u).
\end{equation}
Finally, we bound the last two terms using the entropy inequality:
\begin{equation}
-\ln |\X| \le g_v - \sum_{x \in \X} G(x|u) \ln G(x|u) \le \ln |\X|,
\end{equation}
which leads to the worst-case lower bound:
\begin{equation}
\MMC^* \ge \max_{x} f(x) - \min_{x} f(x) - \ln |\X|.
\end{equation}
\qed

\begin{figure*}
    \centering
    \includegraphics[trim = {0 8cm 0 0}, width=0.9\linewidth]{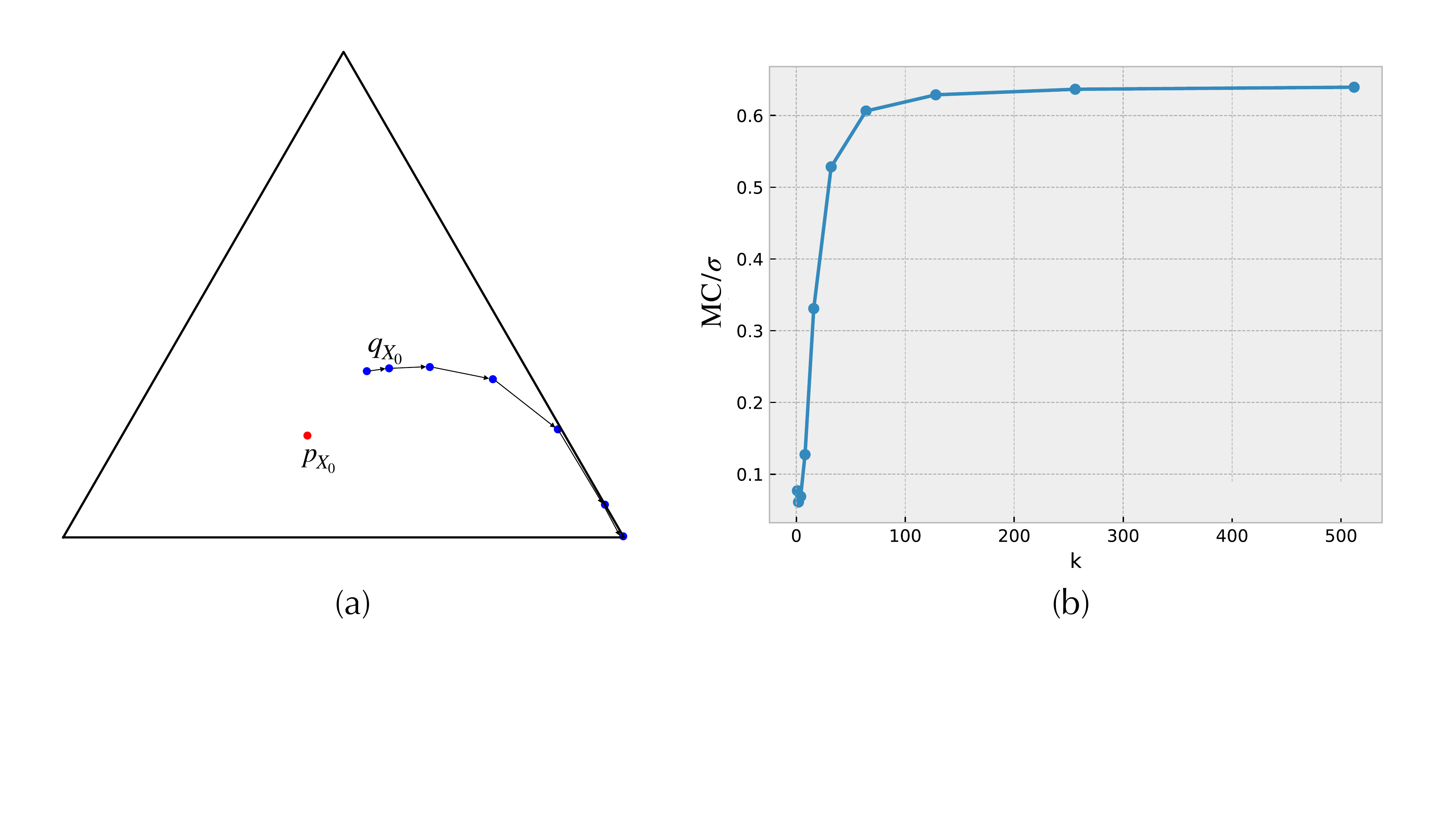}
    \caption{
    Large mismatch cost contribution in the high-EP regime: Starting with arbitrary chosen values $f(1) = 0.5$, $ f(2) = 0.1 $, and $ f(3) = 0.2 $, we scale all $ f(x) $ by a factor $k$, resulting in new values  $k f(x)$.  
    (a) As $k$ increases, the associated prior distribution $q_{X_0} = \mathrm{arg}\min_{r_{X_0}} \C(r_{X_0})$ (as defined by Eq.~\ref{eq:b1}) shifts closer to the boundary of the 2-simplex.  
    (b) Consequently, the mismatch cost for a typical initial distribution (here, $ p_{X_0} = \{0.46, 0.33, 0.21\} $) becomes a larger fraction of the total EP, eventually exceeding $60\%$ of the total EP. The calculations are performed using the map $G$ defined in Eq.~\ref{eq:mapG}, with $\phi$ set to 0.1. The total entropy production is computed from $f(x)$ and $f(x)$ using Eq.~\ref{eq:b1}, while the corresponding mismatch cost for each resulting prior is calculated using $\MMC(p_{X_0}) = D(p_{X_0} \| q_{X_0}) - D( Gp_{X_0} \| Gq_{X_0})$.}
    \label{fig:simplex_MMC}
\end{figure*}

To illustrate this point, consider Fig.~\ref{fig:simplex_MMC}, which shows how the mismatch cost (MMC) contribution to total EP evolves as the function $f(x)$ is scaled by a factor $k$. We use the following stochastic map $G$:

\begin{equation}\label{eq:mapG}
    G = \begin{bmatrix}
        1-\phi & \phi & 0 \\
        0 & 1-\phi & \phi \\ 
        \phi& 0 & 1-\phi
        \end{bmatrix}.
\end{equation}
where $\phi \in [0, 1]$. Starting with small, arbitrary values of $f(x)$, increasing the scaling factor $k$ causes the prior distribution induced by $kf(x)$ to shift toward a corner of the 2-simplex. This shift reflects increasingly peaked priors. Correspondingly, for a typical choice of initial distribution, the MMC becomes a progressively dominant component of the total EP, as demonstrated in Fig.~\ref{fig:simplex_MMC}. 

Next, we examine how the MMC behaves under time coarse-graining. This analysis is crucial because MMC lower bounds can be established at various levels of temporal resolution. To take an example of Boolean circuits, MMC can be evaluated at the fine-grained level where each gate updates sequentially. Alternatively, a coarser time resolution may be adopted, in which groups of gates---such as layers in the circuit---are updated in parallel. At the coarsest level, the analysis considers only the initial and final configurations of the circuit. 

\subsection{Proof that prior is always in the interior of the simplex}\label{app: prior_proof}

\global\long\def\i{x}%
\global\long\def\j{y}%
\global\long\def\Tji{\map(y\vert x)}%
\global\long\def\p{p}%
\global\long\def\s{\mu}%
\global\long\def\N{\mathbb{N}}%
\global\long\def\q{q}%
\global\long\def\N{\mathbb{N}}%
\global\long\def\sX{\X}%
\global\long\def\sY{\mathcal{Y}}%
\global\long\def\hlf{{\textstyle \frac{1}{2}}}%
\global\long\def\e{\epsilon}%
\global\long\def\qip{[\M\q]_{j}}%
\global\long\def\pip{[\M\p](\j)}%
\global\long\def\f{f}%
\global\long\def\E{\mathbb{E}}%
\global\long\def\w{q^{\e}}%
\global\long\def\wi{\w_{i}}%
\global\long\def\wip{\left[\M\w\right]_{j}}%
\global\long\def\wp{\M\w}%
\global\long\def\de{{\textstyle \partial_{\e}^{+}}}%
\global\long\def\dem{{\textstyle \partial_{\e}^{-}}}%
\global\long\def\ppi{p(\i)}%
\global\long\def\ppl{p(c)}%
\global\long\def\ppll{p^{c}}%
\global\long\def\ppil{\ppll(\i)}%
\global\long\def\qe{q^{\e}}%
\global\long\def\pip{Pp(\j)}%
\global\long\def\qip{q(\j)}%
\global\long\def\qll{q^{c}}%

\global\long\def\bj{b(\j)}%
\global\long\def\bi{b(\i)}%
\global\long\def\aj{a(\j)}%
\global\long\def\ai{a(\i)}%

\global\long\def\aej{[\map\ae]_{j}}%
\global\long\def\aei{\ae_{i}}%
\global\long\def\aej{[\map\ae](\j)}%
\global\long\def\aei{\ae(\i)}%
\global\long\def\aej{\ae(\j)}%
\global\long\def\aei{\ae(\i)}%
\global\long\def\ae{a^{\e}}%

Lemma A1 in the extended proofs below establishes that the optimal prior always lies in the interior of the simplex. The proof is adapted from~\cite{kolchinsky2020thermodynamic}, with minor corrections to fix typographical errors present in the original. For the reader’s convenience, we reproduce the corrected version in full.

Consider a conditional distribution $\map(y\vert x)$ that specifies
the probability of ``output'' $y\in\sY$ given ``input'' $x\in\sX$,
where $\sX$ and $\sY$ are finite.

Given some $\sZ\subseteq \sX$, the island decomposition $L_\sZ(P)$ of $\map$, and any $p\in\Delta_\sX$, let $p(c)=\sum_{x\in c}p(x)$ indicate the total probability within
island $c$, and 
\begin{equation}
\ppil:=\begin{cases}
\frac{\ppi}{p(c)} & \text{if \ensuremath{x\in c} and \ensuremath{p(c)>0}}\\
0 & \text{otherwise}
\end{cases}
\end{equation}
indicate the conditional probability of state $\i$ within island
$c$.

In our proofs below, we will make use of the notion of \emph{relative
interior.} Given a linear space $V$, the relative interior of a subset $A\subseteq V$ is defined as,

\begin{equation}
\mathrm{relint}\, A := \{x\in A : \forall y \in A,\exists\epsilon>0 \text{ s.t. } x+\e(x-y)\in A\}\,.
\end{equation}

Finally, for any function $g(x)$, we use the notation
\begin{equation}\partial_{x}^{+}g(x)\vert_{x=a}:=\lim_{\delta \rightarrow0^{+}}\frac{1}{\delta}\left(g(a+\delta)-g(a)\right)
\end{equation}
to indicate the right-handed derivative of $g(x)$ at $x=a$. When the condition that $x=a$ is omitted, $a$ is implicitly
assumed to equal $0$, i.e.,
\begin{equation}\partial_{x}^{+}g(x) :=\lim_{\delta \rightarrow0^{+}}\frac{1}{\delta}\left(g(\delta)-g(0)\right)
\end{equation}
\global\long\def\aej{[\map\ae]_{j}}%
\global\long\def\aei{\ae_{i}}%
\global\long\def\aej{[\map\ae](y)}%
\global\long\def\aei{\ae(x)}%
\global\long\def\aej{\ae(y)}%
\global\long\def\aei{\ae(x)}%
\global\long\def\ae{a^{\e}}%

We also adopt the shorthand that
$\ae\coloneqq a+\e(b-a)$, and write $S(\ae) : = S(p(\ae))$,  $P\ae := Pp(\ae)$, and
so  $S(P\ae) = S(Pp(\ae))$.

%\subsubsection*{Proofs}

Given some conditional distribution $\Tji$ and function $\f:\sX\to\mathbb{R}$, we consider the function $\C:\Delta_\sX \rightarrow\mathbb{R}$
as
\begin{equation}
\C(p):=S(\map p)-S(p)+\E_{p}[f]\,.
\end{equation}
Note that $\C$ is continuous on the relative interior of $\Delta_\sX$.

\begin{applemma}
\label{lem:dd}For any $a,b\in\Delta_\sX$, the directional derivative of
$\C$ at $a$ toward $b$ is given by
\begin{equation}
\de\C(a+\e(b-a))\vert_{\e=0}=D(\map b\Vert\map a)-D(b\Vert a)+\C(b)-\C(a).
\end{equation}
\end{applemma}
\begin{proof}
Using the definition of $\C$,  write
\begin{align}
\de\C(\ae)=\de\left[S(\map\ae)-S(\ae)\right]+\de\E_{\ae}[f].\label{eq:ddd-1}
\end{align}
Consider the first term on the RHS,
\begin{align*}
& \de\left[S(\map\ae)-S(\ae)\right] \\
& =-\sum_{y\in\sY}\left[ (\de P\aej)\ln P\aej+\de[\map\ae](y)\right] \\
& \qquad+\sum_{x\in\sX}\left[(\de \ae)\ln\aei+\de\aei\right]\\
 & =-\sum_{y\in\sY}(Pb(y)-Pa(y))\ln P\aej+\sum_{x\in\sX}(\bi\!-\!a(x))\ln\aei
\end{align*}
Evaluated at $\epsilon=0$, the last line can be written as 
\begin{align*}
 & -\sum_{y\in\sY}(Pb(y)-Pa(y))\ln Pa(y)+\sum_{x\in\sX}(b(x)-a(x))\ln a(x)\\
 & \quad=D(\map b\Vert\map a)+S(\map b)-S(\map a)-D(b\Vert a)-S(b)+S(a)
\end{align*}
where we adopt the convention that if $a(x) = 0, b(x) \ne 0$ for some $x$, then this expression means $-\infty$.
We next consider the $\de\E_{\ae}[f]$ term,
\begin{align*}
\de\E_{\ae}[f] &=\de\left[\sum_{x\in\sX}\left(a(x)+\e(b(x)-a(x))\right)f(x)\right]\\
&=\E_{b}[f]-\E_{a}[f]\,.
\end{align*}

Combining the above gives
\begin{align*}
\de\C(\ae)\vert_{\e=0} & =D(\map b\Vert\map a)-D(b\Vert a)+S(\map b)-S(b)\\
& \qquad -(S(\map a)-S(a))+\E_{b}[f]-\E_{a}[f]\\
 & =D(\map b\Vert\map a)-D(b\Vert a)+\C(b)-\C(a).
\end{align*}
\end{proof}

Importantly, \ref{lem:dd} holds even if there are $x$'s for which $a(x) = 0$ but $b(x) \ne 0$, 
in which case the RHS of the equation in the lemma equals $-\infty$. (Similar comments apply
to the results below.)

\begin{theorem}
\label{thm:1}Let $V$ be a convex subset of $\Delta$. Then for any
$q\in\mathrm{arg}\min_{s\in V}\C(s)$ and any $p\in V$,
\begin{equation}
\C(p)-\C(\q)\ge D(p\Vert\q)-D(\map p\Vert\map\q)\,.\label{eq:thm1es}
\end{equation}
Equality holds if $\q$ is in the relative interior of $V$.
\end{theorem}
\begin{proof}
Define the convex mixture $\qe:=q+\e(p-q)$. %  for $\e\in[0,1]$. 
By \ref{lem:dd}, the
directional derivative of $\C$ at $\q$ in the direction $p-\q$
is
\[
\de\C(\qe)\vert_{\e=0}=D(\map p\Vert\map\q)-D(p\Vert\q)+\C(p)-\C(q)\,.
\]
At the same time, $\de\C(\qe)\vert_{\e=0} \ge 0$, since $q$ is a minimizer within a convex set. 
\ref{eq:thm1es} then follows by rearranging.

When $q$ is in the relative interior of $V$, $q-\epsilon(p-q)\in V$
for sufficiently small $\epsilon>0$. Then,
\begin{align*}
0\le & \lim_{\e\rightarrow0^{+}}\frac{1}{\e}\left(\C(q-\epsilon(p-q))-\C(q)\right)\\
= & -\lim_{\e\rightarrow0^{-}}\frac{1}{\e}\left(\C(q+\epsilon(p-q))-\C(q)\right)\\
= & -\lim_{\e\rightarrow0^{+}}\frac{1}{\e}\left(\C(q+\epsilon(p-q))-\C(q)\right)\\
&=-\de\C(\qe)\vert_{\e=0}.
\end{align*}
where in the first inequality comes from the fact that $q$ is a minimizer,
in the second line we change variables as $\e\mapsto-\e$,
and the last line we use the continuity of $\C$ on interior of the simplex.
Combining with the above implies 
\[\de\C(\w)=D(\map p\Vert\map\q)-D(p\Vert\q)+\C(p)-\C(q)=0.\]
\end{proof}

The following result is key. It means that the prior within an island has full support in that island. 

\begin{applemma}
\label{lem:fsupport}
For any $c\in L(P)$ and $\displaystyle q\in\mathrm{arg}\min_{s : \mathrm{supp}\; s \subseteq c}\C(s)$,
\[
\mathrm{supp}\; q=\{ x \in c : f(x) < \infty \}.
\]
\end{applemma}
\begin{proof}
\global\long\def\qi{q(\i)}%
\global\long\def\qj{q(\j)}%
\global\long\def\ii{\hat{\i}}%
\global\long\def\jj{\hat{\j}}%
\global\long\def\wj{\w(\j)}%
\global\long\def\wi{\w(\i)}%
\global\long\def\Ci{f(\i)}%
\global\long\def\Cii{f(\ii)}%
\global\long\def\Tjjii{\map(\jj\vert\ii)}%
\global\long\def\wjj{\w(\jj)}%
\global\long\def\wii{\w(\ii)}%
\global\long\def\Tjii{\map(\j\vert\ii)}%
\global\long\def\qjj{q(\jj)}%

We prove the claim by contradiction. Assume that $q$ is a minimizer with $\mathrm{supp}\; q\subset \{ x \in c : f(x) < \infty \}$. Note there
cannot be any $\i\in\mathrm{supp}\; q$ and $\j\in\sY\setminus\mathrm{supp}\;\map q$
such that $\Tji>0$ (if there were such an $\i,\j$, then $\qj=\sum_{\i'}P(\j\vert\i')q(\i')\ge\Tji\qi>0$,
contradicting the statement that $\j\in\sY\setminus\mathrm{supp}\;\map q$).
Thus, by definition of islands, there must be an $\ii\in 
c \setminus\mathrm{supp}\; q$,
$\jj\in\mathrm{supp}\;\map q$ such that $f(\ii) < \infty$ and $\Tjjii>0$.

Define the delta-function distribution $u(\i):=\delta(\i,\ii)$ and
the convex mixture $\qe(\i)=(1-\e)q(\i)+\e u(\i)$
for $\e\in[0,1]$. We will also use the notation $\qe(\j)=\sum_{\i}\Tji\qi$.

Since $q$ is a minimizer of $\C$, $\partial_{\e}\C(\qe)\vert_{\e=0}\ge0$.
Since $\C$ is convex, the second derivative $\partial_{\e}^{2}\C(\qe)\ge0$
and therefore $\partial_{\e}\C(\qe)\ge0$ for all $\epsilon\ge0$.
Taking $a=\qe$ and $b=u$ in \ref{lem:dd} and rearranging, we then have
\begin{align}
\C(u) & \ge D(u\Vert\qe)-D(\map u\Vert\map\qe)+\C(\qe) \nonumber \\
 & \ge D(u\Vert\qe)-D(\map u\Vert\map\qe)+\C(q),\label{eq:ineqZ} 
\end{align}
where the second inequality uses that $q$ is a minimizer of $\C$. At the same time, 
\begin{align}
& D(u\Vert\qe)-D(\map u\Vert\map\qe) \nonumber \\
& =\sum_{\j}P(\j\vert\ii)\ln\frac{\qe(\j)}{\qe(\ii)\map(\j\vert\ii)}\nonumber \\
 & =\Tjjii\ln\frac{\qe(\jj)}{\e\Tjjii}+\sum_{\j\ne \jj}P(\j\vert\ii)\ln\frac{\qe(\j)}{\e\map(\j\vert\ii)} \nonumber \\
 & \ge\Tjjii\ln\frac{(1-\e)q(\jj)}{\e\Tjjii}+\sum_{\j\ne \jj}P(\j\vert\ii)\ln\frac{\e\map(\j\vert\ii)}{\e\map(\j\vert\ii)} \nonumber \\
 & =\Tjjii\ln\frac{(1-\e)}{\e}\frac{q(\jj)}{\Tjjii} \label{eq:ineqZZZ} ,
\end{align}
where in the second line we've used that $\qe(\ii)=\e$, and in the third that $\qe(\j)=(1-\e)q(\j) + \e \map(\j\vert \ii)$, so $\qe(\j)\ge(1-\e)q(\j)$ and $\qe(\j) \ge \e \map(\j\vert \ii)$.

Note that the RHS of \ref{eq:ineqZZZ} goes to $\infty$ as $\e \to 0$.  Combined with \ref{eq:ineqZ} and that $\C(q)$ is finite implies that $\C(u) = \infty$.  However, $\C(u) = S(\map(Y\vert \ii)) + f(\ii) \le |\sY| +f(\ii)$, which is finite.  We thus have a contradiction, so $q$ cannot be the minimizer.
\end{proof}

The following result is also key. Intuitively,   it follows from the fact that the directional derivative of $S(p)$ into the
simplex for any $p$ on the edge of the simplex is negative infinite.

\begin{applemma}
\label{lem:unique}
For any island $c\in L(P)$, $\displaystyle q\in\mathrm{arg}\min_{s : \mathrm{supp}\; s \subseteq c}\C(p)$
is unique.
\end{applemma}
\begin{proof}
Consider any two distributions $\displaystyle p,q\in\mathrm{arg}\min_{s:  \mathrm{supp}\; s \subseteq c}\C(s)$, and let $p' = \map p$, $q'=\map q$. We will prove that $p=q$.

First, note that by \ref{lem:fsupport}, $\mathrm{supp}\; q = \mathrm{supp}\; p = c$.  By \ref{thm:1},
\begin{align*}
\C(p)-\C(q)&=D(p \Vert q)-D(p' \Vert q') \\
& = \sum_{\i,\j} p(x) \Tji \ln \frac{p(\i) q'(\j)}{q(\i)p'(\j)} \\
& = \sum_{\i,\j} p(x) \Tji \ln \frac{p(\i) \Tji}{q(\i)p'(\j) \Tji / q'(\j)} \\
& \ge 0
\end{align*}
where the last line uses the log-sum inequality.  If the inequality is strict, then  $p$ and $q$ can't 
both be minimizers, i.e., the minimizer must be unique, as claimed. 

If instead the inequality is not strict, i.e., $\C(p)-\C(q) = 0$, then there is some constant $\alpha$ such that for all $\i,\j$ with $\Tji > 0$,
\begin{align}
 \frac{p(\i) \Tji}{q(\i)p'(\j) \Tji / q'(\j)}=\alpha  
\end{align}
which is the same as
\begin{align}
 \frac{p(\i)}{q(\i)} = \alpha \frac{p'(\j)}{q'(\j)}.
 \label{eq:lsecond}
\end{align}

Now consider any two different states $x,x'\in c$ such that
$\map(y\vert x)>0$ and $\map(y\vert x')>0$ for some $y$ (such states
must exist by the definition of islands). For \ref{eq:lsecond} to hold for both $x, x'$ with that same, shared $y$, it must be that 
${p(x)}/{q(x)}={p(x')}/{q(x')}$. Take another state
$x''\in c$ such that $\map(y'\vert x'')>0$ and $\map(y'\vert x')>0$
for some $y'$. Since this must be true for all pairs $x, x' \in c$, ${p(x)}/{q(x)}=\text{const}$ for all $x\in c$, and $p=q$, as claimed.
\end{proof}

\begin{applemma}
\label{thm:decomp}$\C(p)=\sum_{c\in L(P)}p(c)\C(\ppll)$.
\end{applemma}
\begin{proof}
First, for any island $c\in L(P)$, define 
\[
\phi(c)=\{\j\in\sY:\exists\i\in c\text{ s.t. }\Tji>0\}\,.
\]
In words, $\phi(c)$ is the subset of output states in $\sY$ that
receive probability from input states in $c$. By the definition
of the island decomposition, for any $\j\in\phi(c)$, $\Tji>0$ only
if $\j\in c$. Thus, for any $p$ and any $\j\in\phi(c)$, we can
write
\begin{equation}
\frac{\pip}{p(c)}=\frac{\sum_{\i}\Tji\ppi}{p(c)}=\sum_{\i\in\sX}\Tji\ppil\,.\label{eq:id0}
\end{equation}

Using $p(x)=\sum_{c\in L(P)}p(c)\ppll(x)$
and linearity of expectation, write $\E_{p}[f]=\sum_{c\in L(P)}p(c)\E_{\ppll}[f]$.

Then,
\begin{align*}
& S(\map p)-S(p) \\
& =-\sum_{\j}\pip\ln\pip+\sum_{\i}\ppi\ln\ppi\\
 & =\sum_{c\in L(P)}p(c)\Big[-\sum_{\j\in\phi(c)}\frac{\pip}{p(c)}\ln\frac{\pip}{p(c)}+\sum_{\i\in c}\frac{\ppi}{p(c)}\ln\frac{\ppi}{p(c)}\Big]\nonumber \\
 & =\sum_{c\in L(P)}p(c)\left[S(\map\ppll)-S(\ppll)\right],
\end{align*}
where in the last line we've used \ref{eq:id0}. Combining gives
\begin{align*}
\C(p)&=\sum_{c\in L(P)}p(c)\left[S(\map\ppll)-S(\ppll)+\E_{\ppll}[f]\right]\\
& =\sum_{c\in L(P)}p(c)\C(\ppll)\,.
\end{align*}
\end{proof}

We are now ready to prove the main result of this appendix.

\begin{theorem}
\label{thm:cost}Consider any function $\C :\Delta_\sX \to \mathbb{R}$ of the form
\begin{align*}
\C(p) := S(\map p) - S(p) + \mathbb{E}_p[f]  
\end{align*}
where  $\map(y\vert x)$ is some conditional distribution of $y \in \sY$ given
 $x\in \sX$ and $f : \sX \to \mathbb{R} \cup \{ \infty \}$ is some function. Let $\sZ$ be any subset of $\sX$ such that $f(x)<\infty$ for $x\in\sZ$, and let $q \in \Delta_\sZ$ be any distribution that obeys 
\[
q^c \in \mathrm{arg}\min_{{\rr : \mathrm{supp}\; \rr \subseteq c}}  \C(\rr)  \;\;\text{ for all }\;\; c \in L_\sZ(\map).
\]
Then, each $q^c$ will be unique, and for any $p$ with $\mathrm{supp}\; p \subseteq \sZ$,
\begin{align*}
\C(p) & = \DDf{p}{q} - \DDf{\map p}{\map q} + \sum_{\mathclap{c \in L_\sZ(\map)}} p(c) \C(q^c)  .
\end{align*}
\end{theorem}
\begin{proof}
We prove the theorem by considering two cases separately.
\vspace{5pt}

\noindent \textbf{Case 1}: $\sZ = \sX$. This case can be assumed when $f(x)<\infty$ for all $x$, so that $L_\sZ(\map) = L(\map)$.  
Then, by \ref{thm:decomp}, we have $\C(p)=\sum_{c\in L(P)}p(c)\C(\ppll)$.
By \ref{lem:fsupport} and \ref{thm:1},
\[
\C(\ppll)-\C(\qll)=D(\ppll\Vert\qll)-D(\map\ppll\Vert\map\qll),
\]
where we've used that if some $\mathrm{supp}\;\qll=c$, then $\qll$ is in
the relative interior of the set $\{ s \in \Delta_\sX : \mathrm{supp}\; s \subseteq c \}$. $\qll$ is unique by \ref{lem:unique}.

At the same time, observe
that for any $p,r\in\Delta_\sX$,
\global\long\def\ri{r(\i)}%
\global\long\def\rj{Pr(y)}%
\begin{align*}
& D(p\Vert r)-D(\map p\Vert\map r)\\
& = \sum_{\i}\ppi\ln\frac{\ppi}{\ri}-\sum_{\j}\pip\ln\frac{\pip}{\rj}\\
&= \sum_{c\in L(P)}p(c)\Bigg[\sum_{\i\in c}\frac{\ppi}{p(c)}\ln\frac{\ppi/p(c)}{\ri/r(c)} \\
& \qquad\qquad\qquad\qquad-\sum_{\j\in\phi(c)}\frac{\pip}{p(c)}\ln\frac{\pip/p(c)}{\rj/r(c)}\Bigg]\\
 &= \sum_{c\in L(P)}p(c)\left[D(\ppll\Vert r^{c})-D(\map\ppll\Vert\map r^{c})\right]\,.
\end{align*}
The theorem follows by combining.

\vspace{5pt}

\noindent \textbf{Case 2}: $\sZ \subset \sX$. In this case, define a ``restriction'' of $f$ and $\map$ to domain $\sZ$ as follows:
\begin{enumerate}
\item Define $\tilde{f} : \sZ \to \mathbb{R}$ via $\tilde{f}(\i)=f(x)$ for $\i\in \sZ$. %ote that by assumption, $\tilde{f} : \sZ \to \mathbb{R}$. 
\item Define the conditional distribution $\tilde{\map}(\j\vert\i)$ for $\j\in \sY,\i \in \sZ$ via $\tilde{\map}(\j\vert\i)=\Tji$ for all $y\in\sY,\i\in \sZ$. 
\end{enumerate}
In addition, for any distribution $p \in \Delta_\sX$ with $\mathrm{supp}\; p \subseteq \sZ$, let $\tilde{p}$ be a distribution over $\sZ$ defined via $\tilde{p}(\i)=p(\i)$ for $\i \in \sZ$.  Now, by inspection, it can be verified that for any  $p \in \Delta_\sX$ with $\mathrm{supp}\; p \subseteq \sZ$,
\begin{align}
\C(p) = S(\tilde{\map}\tilde{p})-S(\tilde{p})+\E_{\tilde{p}}[\tilde{f}] =: \tilde{\C}(\tilde{p})
\label{eq:www99}
\end{align}
We can now apply \text{Case 1} of the theorem to the function $\tilde{\C}:\Delta_\sZ\to\mathbb{R}$, as  defined in terms of the tuple $({\sZ}, \tilde{f}, \tilde{\map})$ (rather than the function $\C:\Delta_\sX\to\mathbb{R}$, as defined in terms of the tuple 
$(\sX, {f}, {\map})$).  This gives 
\begin{align}
\tilde{\C}(\tilde{p}) = D(\tilde{p}\Vert \tilde{q}) -  D(\tilde{\map}\tilde{p}\Vert \tilde{\map} \tilde{q}) + \sum_{c \in \LLL(\tilde{\map})} \tilde{p}(c) \tilde{\C}(\tilde{q}^c),
\label{eq:B1}
\end{align}
where, for all $c \in \LLL(\tilde{\map})$, $\tilde{q}^c$ is the unique distribution that satisfies
$\tilde{q}^c \in \mathrm{arg}\min_{r \in \Delta_\sZ :\mathrm{supp}\; r \subseteq c} \tilde{\C}(r)$.

Now, let $q$ be the natural extension of $\tilde{q}$ from $\Delta_\sZ$ to $\Delta_\sX$.  Clearly,  for all $c \in \LLL(\tilde{\map})$, ${\C}({q}^c)=\tilde{\C}(\tilde{q}^c)$ by \ref{eq:www99}. In addition, each $q^c$  is the unique distribution that satisfies 
${q}^c \in \mathrm{arg}\min_{r \in \Delta_\sX :\mathrm{supp}\; r \subseteq c} {\C}(r)$. 
Finally, it is easy to verify that $D(\tilde{p}\Vert \tilde{q}) = D({p}\Vert {q})$, $D(\tilde{\map}\tilde{p}\Vert \tilde{\map} \tilde{q}) = D({\map}{p}\Vert {\map} {q})$, $L(\tilde{\map}) = L_\sZ(\map)$.  Combining the above results with \ref{eq:www99} gives
 \[
 \C(p) = \tilde{\C}(\tilde{p}) = D({p}\Vert {q}) -  D({\map}{p}\Vert {\map} {q}) + \sum_{\mathclap{c \in \LLL_\sZ({\map})}} {p}(c) {\C}({q}^c).
 \]
\end{proof}

\end{document}